\documentclass[fleqn,usenatbib]{mnras}

\usepackage{newtxtext,newtxmath}
\usepackage{pdflscape}

\usepackage[T1]{fontenc}

\DeclareRobustCommand{\VAN}[3]{#2}
\let\VANthebibliography\thebibliography
\def\thebibliography{\DeclareRobustCommand{\VAN}[3]{##3}\VANthebibliography}

\usepackage{graphicx}	
\usepackage{amsmath}	
\usepackage{longtable}
\usepackage{ulem}

\newcommand{\ignore}[1]{}
\newcommand{\yc}{\mbox{$\tilde{y}_0$}}
\newcommand{\ym}{$Y\!\text{-}M$}

\title[Compact sources in galaxy clusters]{Observations of compact sources in galaxy clusters using MUSTANG2}

\author[S. R. Dicker et al.]{
Simon R. Dicker,$^{1}$ Elia S. Battistelli,$^{2}$ Tanay Bhandarkar,$^{1}$ Mark J. Devlin,$^{1}$ Shannon M. Duff,$^{3}$
\newauthor Gene Hilton ,$^{3}$ Matt Hilton,$^{4,5}$ Adam D. Hincks,$^{6}$ Johannes Hubmayr,$^{3}$ Kevin Huffenberger,$^{7}$
\newauthor John P. Hughes,$^{8}$ Luca Di Mascolo,$^{9,10,11}$ Brian S. Mason,$^{12}$ J.A.B. Mates,$^{3}$ Jeff McMahon,$^{13,14,15,16}$ 
\newauthor Tony Mroczkowski,$^{17}$ Sigurd Naess,$^{18}$ John Orlowski-Scherer,$^{1}$  Bruce Partridge,$^{19}$ 
\newauthor Federico Radiconi,$^{2}$ Charles Romero,$^{20,1}$ Craig L. Sarazin,$^{21}$ Neelima Sehgal,$^{22}$ Jonathan Sievers,$^{23}$ 
\newauthor Crist\'obal Sif\'on,$^{24}$ Joel Ullom,$^{3}$ Leila R. Vale,$^{3}$ Michael R. Vissers,$^{3}$ Zhilei Xu,$^{25,1}$ 
\\
$^{1}$Department of Physics and Astronomy, University of Pennsylvania, 209 South 33rd Street, Philadelphia, PA, 19104, USA\\
$^{2}$Sapienza University of Rome,Physics Department, Piazzale Aldo Moro 5, 00185 Rome, Italy\\
$^{3}$NIST, Quantum Sensors Group, 325 Broadway, Boulder, CO, 80305, USA\\
$^{4}$Astrophysics Research Centre, University of KwaZulu-Natal, Westville Campus, Durban 4041, South Africa\\
$^{5}$School of Mathematics, Statistics \& Computer Science, University of KwaZulu-Natal, Westville Campus, Durban 4041, South Africa\\
$^{6}$David A. Dunlap Department of Astronomy \& Astrophysics, University of Toronto, 50 St. George St., Toronto ON M5S 3H4, Canada\\
$^{7}$Department of Physics, Florida State University, Tallahassee, Florida 32306 USA\\
$^{8}$Department of Physics and Astronomy, Rutgers, the State University of New Jersey, 136 Frelinghuysen Road, Piscataway, NJ 08854-8019, USA\\
$^{9}$Astronomy Unit, Department of Physics, University of Trieste, via Tiepolo 11, Trieste 34131, Italy\\
$^{10}$INAF - Osservatorio Astronomico di Trieste, via Tiepolo 11, Trieste 34131, Italy\\
$^{11}$IFPU - Institute for Fundamental Physics of the Universe, Via Beirut 2, 34014 Trieste, Italy\\
$^{12}$National Radio Astronomy Observatory, 520 Edgemont Rd., Charlottesville VA 22903, USA\\
$^{13}$Department of Astronomy and Astrophysics, University of Chicago, 5640 S. Ellis Ave., Chicago, IL 60637, USA\\
$^{14}$Kavli Institute for Cosmological Physics, University of Chicago, 5640 S. Ellis Ave., Chicago, IL 60637, USA \\
$^{15}$Department of Physics, University of Chicago, Chicago, IL 60637, USA\\
$^{16}$Enrico Fermi Institute, University of Chicago, Chicago, IL 60637, USA\\
$^{17}$European Southern Observatory, Karl-Schwarzschild-Strasse 2, Garching D-85748, Germany\\
$^{18}$Center for Computational Astrophysics, Flatiron Institute, New York, NY, 10010, USA\\
$^{19}$Department of Astronomy, Haverford College, 370 Lancaster Ave, Haverford PA 19041, USA\\
$^{20}$Green Bank Observatory, 155 Observatory Road, Green Bank, WV 24944. USA\\
$^{21}$Department of Astronomy, University of Virginia, 530 McCormick Road, Charlottesville, VA 22904-4325, USA\\
$^{22}$Physics and Astronomy Department, Stony Brook University, Stony Brook, NY 11794, USA\\
$^{23}$Department of Physics, McGill University, 3600 University Street Montreal, QC H3A 2T8, Canada\\
$^{24}$Instituto de F\'isica, Pontificia Universidad Cat\'olica de Valpara\'iso, Casilla 4059, Valpara\'iso, Chile\\
$^{25}$MIT Kavli Institute, Massachusetts Institute of Technology, 77 Massachusetts Avenue, Cambridge, MA 02139, USA\\}

\date{Accepted 22/9/2021}

\pubyear{2021}

\begin{document}
\label{firstpage}
\pagerange{\pageref{firstpage}--\pageref{lastpage}}
\maketitle

\begin{abstract}
Compact sources can cause scatter in the scaling relationships between the amplitude of the thermal Sunyaev-Zel'dovich Effect (tSZE) in galaxy clusters and cluster mass.  Estimates of the importance of this scatter vary -- largely due to limited data on sources in clusters at the frequencies at which tSZE cluster surveys operate.  In this paper we present 90~GHz compact source measurements from a sample of 30 clusters observed using the MUSTANG2 instrument on the Green Bank Telescope. We present simulations of how a source’s flux density, spectral index, and angular separation from the cluster’s center affect the measured tSZE in clusters detected by the Atacama Cosmology Telescope (ACT). By comparing the MUSTANG2 measurements with these simulations we calibrate an empirical relationship between 1.4~GHz flux densities from radio surveys and  source contamination in ACT tSZE measurements.  We find  3 per cent  of the ACT clusters have more than a 20 per cent  decrease in Compton-$y$ but another 3 per cent  have a 10 per cent {\it increase} in the Compton-$y$ due to the matched filters used to find clusters. As sources affect the measured tSZE signal and hence the likelihood that a cluster will be detected, testing the level of source contamination in the tSZE signal using a tSZE selected catalog is inherently biased.  We confirm this by comparing the ACT tSZE catalog with optically and X-ray selected cluster catalogs.  There is a strong case for a large, high resolution survey of clusters to better characterize their source population. 
\end{abstract}
 



\section{Introduction}
Galaxy clusters are important probes of cosmology and are laboratories for the study of the highest energy events since the Big Bang.  Consequently, much effort has gone into surveys to find them.  The first surveys \citep{Abell1958,Zwicky1961} used over-densities of galaxies to locate clusters, but with the dawn of X-ray astronomy, in the late 1960's, searches for clusters relying on the emission from the hot intracluster medium (ICM) became possible \citep[e.g.][]{Uhuru,Einstein_clusters,x-ray_survey}. 
In the last two decades, the thermal Sunyaev-Zel'dovich effect \citep[tSZE;][]{Sunyaev1972} has been added to the toolkit to both find clusters and study their ICM \citep[see e.g.][for a review]{Mroczkowski2019}. 
Rather than emission, the tSZE consists of the inverse Compton scattering of cosmic microwave background (CMB) photons as they pass through the ICM which causes a spectral distortion of the  CMB blackbody. The magnitude of the effect in any one direction is proportional to the pressure integrated along the line of sight and is referred to as Compton-$y$. The total Compton-$y$ integrated across the angular extent of the cluster, $Y$, is a good proxy for cluster mass \citep[e.g. ][]{Kravtsov2012}. 

One of the advantages of the tSZE is that its surface brightness is redshift independent, meaning that, given sufficient resolution, it is relatively easy to detect and study clusters at high redshifts where optical and X-ray methods need long exposure times.  Experiments such as \textit{Planck} \citep{PlanckInstrument} and the South Pole Telescope (SPT) \cite{Benson2014} have carried out deep surveys finding hundreds of clusters \citep{Planck_SZE,SPT_SZE}  
and the recent data release from the Atacama Cosmology Telescope \citep[ACT;][]{ACT} contains 4195 optically confirmed clusters \citep{Hilton2021}. In the future, experiments such as the Simons Observatory, CMB-S4, 
and CMB-HD expect to find an order of magnitude more clusters \citep{SimonsForecastPaper, abazajian2016cmbs4, Abazajian2019, CMB_HD}. Because of the large size of current and future tSZE surveys, their use for cosmology will be limited by systematic effects.  The selection function of these surveys is relatively well understood -- due to their insensitivity to redshift, it is mostly a selection by mass -- however, as we will show later, one possible systematic effect is the exclusion of some clusters due to radio sources. Although relationships between $Y$ and cluster mass exist, there is an 11 per cent  scatter, with some clusters deviating from the relationship by up to 15 per cent  \citep{YMRelationship}.  A good understanding of the causes of this scatter is essential to realizing the full potential of future data sets.

\begin{figure}
    \centering
	\includegraphics[clip,trim=5mm 2mm 5mm 0mm, width=0.49\columnwidth]{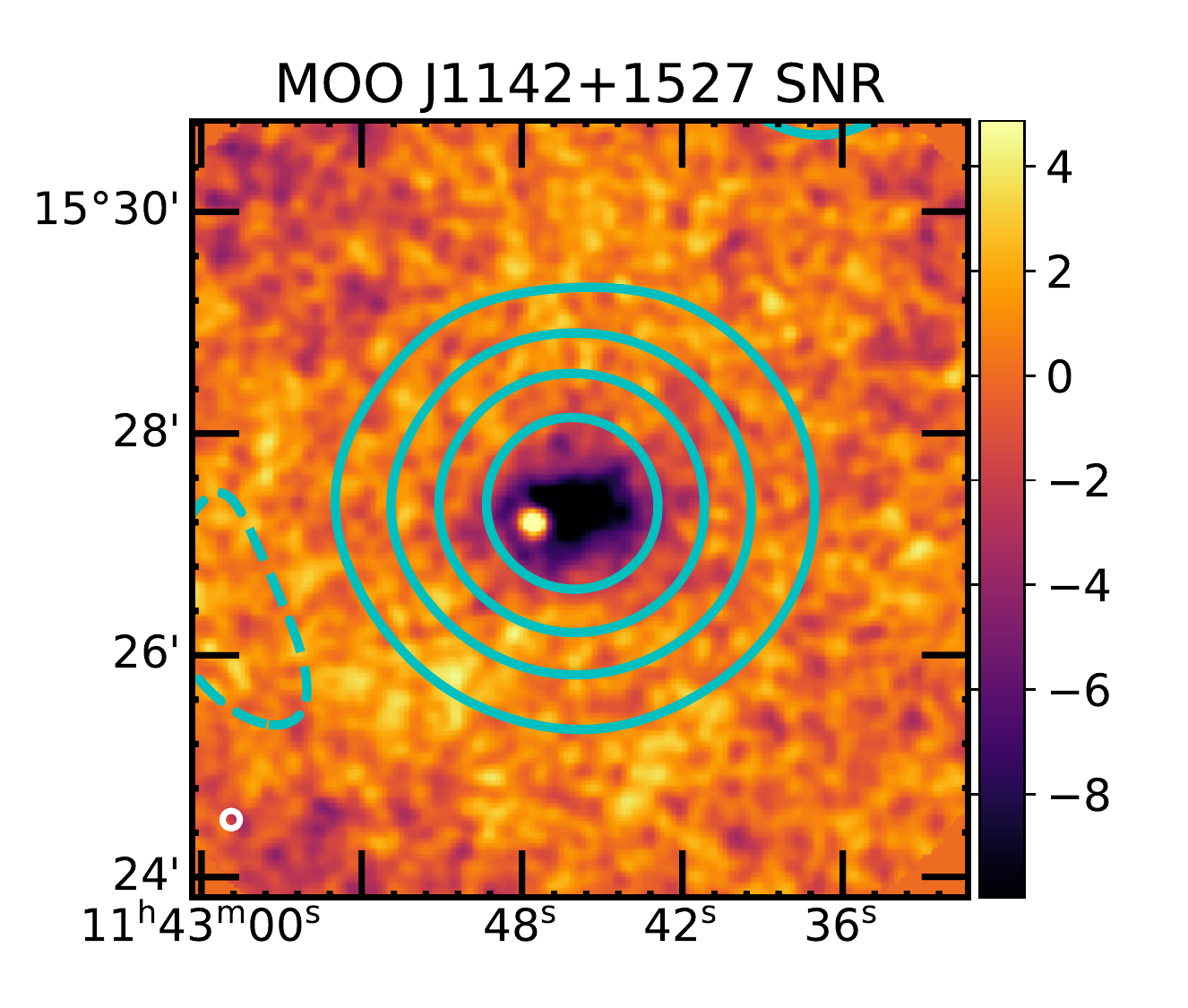}
	\includegraphics[clip,trim=5mm 2mm 5mm 0mm, width=0.49\columnwidth]{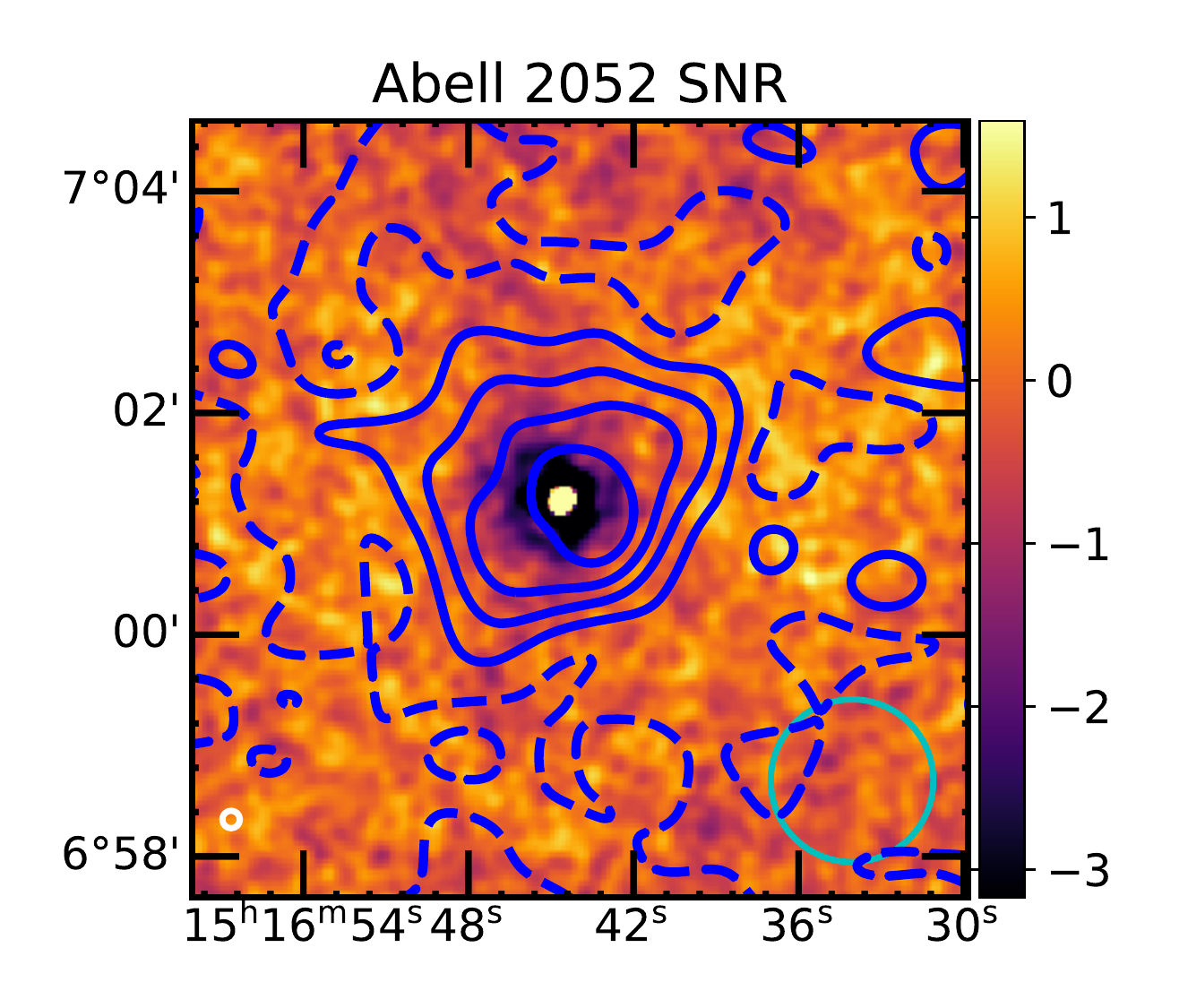}
    \caption{\label{fig:maps} MUSTANG2 signal-to-noise ratio maps of two clusters.  In MOO J1142+1527; z=1.189 (left), the filtered ACT $y$ map is shown as cyan contours spaced by $5{\times}10^{-5}$. The bright point source  and the elliptical center of this on-going merger do not show up in the ACT $y$ map. In Abell 2052; z=0.03 (right) the blue contours are the ACT 90~GHz map with 100~$\mu$K spacing.  The ACT beam is shown in cyan.  The point source in this cluster is strong enough that all ACT sees is the point source and this well known cluster is missing from the ACT DR5 sample.  The MUSTANG2 beam is shown (in white) in the bottom left of both maps.  
    }
\end{figure}

In order to maximize their survey speed, experiments such as ACT and the SPT have moderate resolutions (1--2 arcmin) which is well matched to the typical angular size of a cluster.  As Fig.~\ref{fig:maps} shows, this means they are unable to resolve features in the ICM such as elliptical cores or shock fronts. Such features can indicate  which clusters are undergoing mergers -- events that can affect the \ym\ relationship \citep{MergersAndSZE}.  Also visible in these 9~arcsec resolution maps, taken with MUSTANG2 \citep{Dicker2014}, are point sources. At 1--2 arcmin resolution these blend in with the clusters contributing to the scatter in the measured $Y$. If this source population is well quantified it can be taken into account when fitting cluster mass to the \ym\ relation.  The measured masses of individual clusters will vary from their true values but, when taken as a whole, the survey will give the correct distribution of masses. Many studies have predicted source contamination levels derived from source catalogs at frequencies well below and well above the tSZE bands \citep[e.g.,][]{knox2004,Lin2007,SPT_src,Lin2009}.  However, there can be large uncertainties in the spectral indices used to extrapolate to tSZE frequencies and not all studies agree. The best solution is to simply measure source properties at the frequencies of interest (90--150~GHz for current surveys).
In this paper we present a pilot study using data from  MUSTANG2 to better quantify the effect of point sources on tSZE derived cluster masses. 

This paper is organized as follows: In Section~\ref{sec:M2obs} we describe the MUSTANG2 observations. Next, in Section~\ref{sec:ACT}, we give an outline of how clusters are found in tSZE surveys using the example of ACT and present simulations showing how point sources will affect the measurements. In Section~\ref{sec:results} we apply the results of these simulations to different samples of clusters, we use extrapolations from low frequency data to compare tSZE, optical, and X-ray selected samples and we  calibrate these extrapolations using the point sources in clusters observed by MUSTANG2.  The conclusions are presented in Section~\ref{sec:conclusions}. We assume a flat cosmology with $\Omega_\text{m} = 0.3$, $\Omega_\Lambda = 0.7$, and $H_0 = 70~\text{km}\,\text{s}^{-1}$ throughout this paper. 

\section{MUSTANG2 observations of clusters}\label{sec:M2obs}
MUSTANG2 is a 90~GHz bolometer camera on the 100~m Green Bank Telescope \citep{Dicker2014}. It has 9 arcsec resolution, a 4.2 arcmin field-of-view and can map a 6~arcmin diameter area to 56~$\mu$Jy/beam in an hour making it ideal for follow up observations of galaxy clusters.
As part of a wider observing program with goals ranging from solar system and galactic science 
to cosmology, MUSTANG2 has mapped over 40 galaxy clusters with typical map depths of 15--50~$\mu \mbox{K}_{\mbox{\tiny RJ}} $ (11--38~$\mu$Jy\,beam$^{-1}$ or
3--10~$\mu\mbox{K}_{\mbox{\tiny CMB}}$\,arcmin). Science goals of these cluster observations range from searching for substructure such as bubbles and shocks, measuring ICM profiles
\citep{Romero2020}, looking for filaments between clusters \citep{Adam2021}, and the follow-up of clusters identified by surveys such as the Massive and Distant Clusters of WISE Survey \citep[MaDCoWS - ][]{Gonzalez2019} and Hyper Suprime-Cam \citep[][]{Okabe2021}.  These observations are spread over many different projects and on their own they do not give any statistically significant information about the population of point sources in clusters. However, by combining all public data we are able to construct a sample of 30 clusters which have a clear detection of the tSZE.  These clusters span redshifts between 0.03 and 1.8 but most are above $z=0.4$ and the sample has a median redshift of 1.035.  Included in the sample are well known clusters such as RXJ1347.5-1145 \citep{Mason2010} and MACSJ0717.5+3745 \citep{Mroczkowski2012}, tSZE identified clusters from ACT \citep{Hasselfield2013}, and 20 clusters from ongoing follow up of clusters identified by MaDCoWS \citep{Dicker2020}. All but two of the clusters are either optically or X-ray selected. 

To extract the point sources, MUSTANG2's MIDAS pipeline \citep[see][for details]{Romero2020} was used to calibrate the raw data and produce signal and noise maps with 1 arcsec pixel spacing.  The noise maps were made by inverting half of each cluster's data.   For the most conservative numbers, the first half of each night's observations was subtracted from the second half (thus, long timescale drifts that are not removed in data analysis are included in the noise estimates). From these maps, signal-to-noise (SNR) maps smoothed to 9 arcsec were produced and the approximate locations of any sources more than 4.5 sigma above the average value of the surrounding pixels  and within 5~arcmin of the center of the cluster where found.  A non-linear least squares fit of a 2D Gaussian over a 20 arcsec square region around each source was used to find the source sizes and peak amplitudes (allowing for a 2D linear offset over the 20 arcsec region).  Where the source size was statistically larger than 10~arcsec it was assumed to be extended and an integrated flux was calculated using the ratio of the source and beam solid angles. For spectral index calculations the integrated fluxes were used whenever a source was extended.

As well as flux densities for sources visible in the maps, we place limits on those that might not have been detected.  The noise in Jy\,beam$^{-1}$ can be calculated by smoothing the noise maps and taking the RMS over the central arcminute, a spatial scale on which the noise is relatively white. From the noise in each map the 90 per cent  completeness limits for point sources in Table~\ref{tab:clusters} are calculated -- in all but 2 of the clusters, the 90 per cent  completeness limit was better than 0.2~mJy.  Since the coverage in the MUSTANG2 maps falls off with radius, these numbers are only applicable to the central $r{\approx}2$~arcmin of the maps. The noise (hence the detection threshold) increases by a factor of $\sim2$ by r=3~arcmin and up to a factor of 3 by r=4~arcmin, but as shown in Section~\ref{sec:sims}, sources at these distances from the cluster centers are correspondingly less important to measurements of Compton-$y$. 

\subsection{MUSTANG2 sources}
\begin{figure}
	\includegraphics[width=\columnwidth]{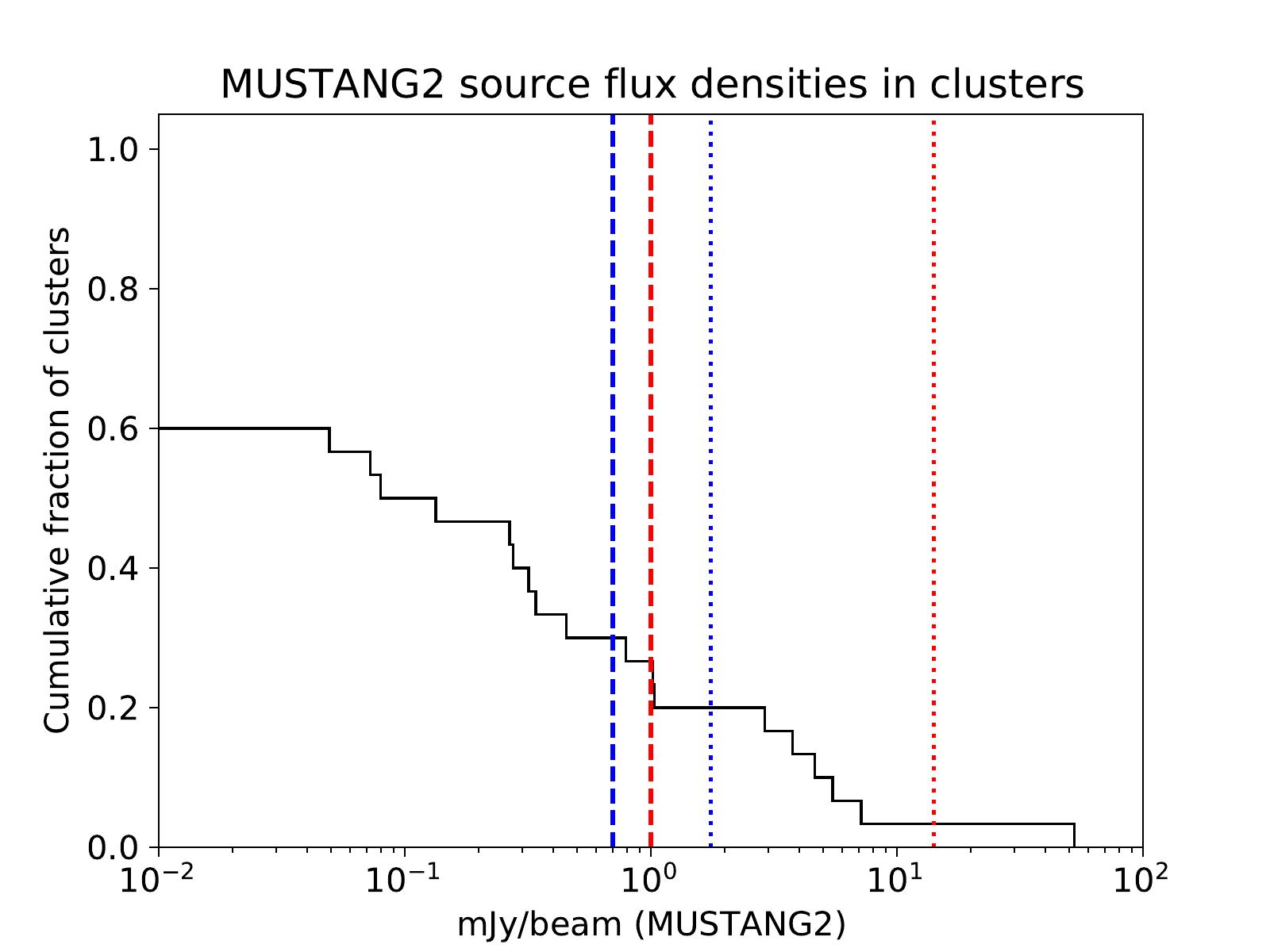}
    \caption{\label{fig:hist} The fraction of clusters that have point source flux densities totaling more than $s$, where $s$ is plotted on the x axis. The dotted vertical lines represent the flux density cutoff used in the ACT point source mask used in DR5 (10~mJy at 150~GHz) scaled to 90~GHz assuming a typical dust spectral index of 3.5 (blue) and a typical synchrotron spectral index of $-0.7$ (red).  The vertical dashed lines represent the flux densities which, {\it if the source were in the center a cluster}, would cause a 5 per cent  reduction in the measured Compton-$y$ for a cluster with $M_{500\text{c}}$=$2.5{\times}10^{14}~\mbox{M}_\odot$\ for the cases of a dusty source (in blue) and a radio source (in red).}
\end{figure}

The locations and flux densities of point sources found  are listed in Table~\ref{tab:src}.  Of our 30 clusters, 18 had one or more sources visible at the depth of the available data.  This is far higher than what would be expected from the chance alignment of foreground and background sources -- extrapolating source counts from the  31~GHz results in \citet{Mason2009} predicts that less than 5 per cent  of clusters should have a source brighter than 1~mJy.  As the cumulative histogram in   Fig.~\ref{fig:hist} shows, 20 per cent  of our cluster sample have sources totalling more than 1~mJy, more than can be explained by any reasonable flattening of spectral indices such as described in \citet{Whittam2017}. This implies most of the sources measured by MUSTANG2 are either cluster members or lensed background sources. 

When searching for clusters, the ACT DR5 pipeline masks out areas of the maps that have point sources with measured amplitudes above 10~mJy at 150~GHz.  For typical radio sources with spectral indices of $-0.7$ this corresponds to 14.2~mJy in MUSTANG2's band.  From Fig.~\ref{fig:hist}, it can be seen that only one of the clusters would have been (and was) masked out. Calculations based on the simulations in Section~\ref{sec:sims} show that an embedded radio source with an amplitude of 0.4~mJy would have a 5 per cent  effect on the Compton-$y$ measured for a $2.5{\times}10^{14}~\text{M}_\odot$\ cluster (approximately the mean mass in the latest ACT cluster catalog). Similarly, dusty sources with spectral indices of 3.5 and amplitudes less than 1.75~mJy at 90~GHz would not be strong enough to be cut by the ACT mask, but a dusty source as faint as 0.07~mJy would still be strong enough to cause a 5 per cent  change in the measured Compton-$y$. From Fig.~\ref{fig:hist} it can be seen that, regardless of the source type, a significant number of the sources found by MUSTANG2 have flux densities that could bias surveys when they lie close to the center of the clusters.

\begin{figure}
	\includegraphics[width=\columnwidth]{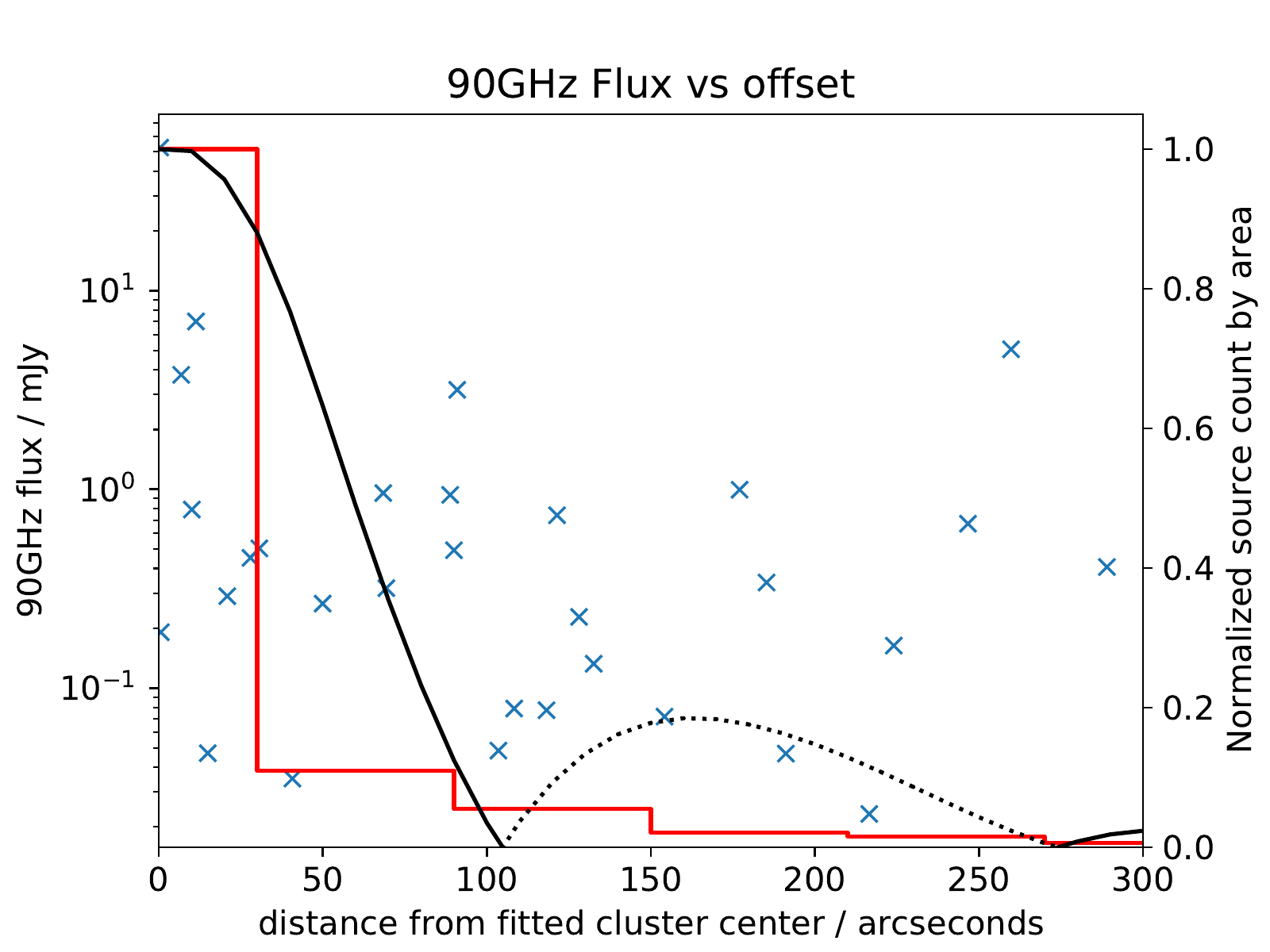}
    \caption{\label{fig:dist} The source flux densities found by MUSTANG2 (the blue crosses) plotted against their distance from the cluster centers. The flux density scale is given on the left. The red histogram shows the distribution of these sources in 1 arcmin bins normalized by area -- although the sources are centrally concentrated, in absolute numbers there are roughly equal numbers of sources at each radius.  The black line shows the absolute value of $N(r)$, the normalized response discussed in section 3.1. The dotted part of this line represents where $N(r)$ is negative.}
\end{figure}

As discussed in Section~\ref{sec:sims} the magnitude of a source's effect on measurements of the tSZE is dependant on its distance from the cluster's center and its spectral index at the frequencies at which the tSZE survey is carried out.  The distance from the cluster center of all the sources found by MUSTANG2 is shown in Fig.~\ref{fig:dist}.  Within the limits of our sample size, the distribution with radius seems uniform and independent of source amplitude. If the source counts are binned in radius and normalized by the solid angle of each bin then it can be seen that the source distribution is centrally peaked. The uniform distribution with radius implies  that the source density  (in projection) must fall by approximately the radius squared. Given the small number statistics due to our sample size, this source distribution is consistent with \citet{SPT_src} who observed a radial distribution of sources of $\sim r^{-3}$ ($r^{-2.5}$ in projection). 

\begin{table}
\caption{The clusters in our sample, their redshifts, and masses (measured from fits to the MUSTANG2 data), and the noise in the centers of the MUSTANG2 maps in $\mu\mbox{K}_{\mbox{\tiny RJ}}$ and also expressed as the 90 per cent  completeness limit, $L90$, for point sources.\label{tab:clusters} }
\begin{tabular}{lcccc}
\hline
Cluster ID & redshift   & $M_{500\text{c}}$ & map noise & $L90$ \\
          &           & $10^{14}\mbox{M}_\odot $ & $\mu\mbox{K}_{\mbox{\tiny{RJ}}}$ & mJy \\ \hline

ACT-CL J0059-0049   & 0.787     & $4.19^{\mbox{\tiny $+0.49-0.62$}}$ &  35  & 0.14 \\
MOO J0105+1323   & 1.130       & $3.83^{\mbox{\tiny $+0.23-0.24$}}$ &  41  & 0.15 \\
MOO J0135+3207   & 1.460     & $1.82^{\mbox{\tiny $+0.31-0.31$}}$ &  29  & 0.10 \\
HSC J0210-0611   & 0.434     & $1.41^{\mbox{\tiny $+0.18-0.23$}}$ &  71  & 0.28 \\
HSC J0221-0346   & 0.430     & $4.41^{\mbox{\tiny $+0.69-1.41$}}$ &  13  & 0.05 \\
HSC J0233-0530   & 0.420     & $1.28^{\mbox{\tiny $+0.31-0.37$}}$ &  21  & 0.08 \\
ACT-CL J0326-0043   & 0.447     & $4.49^{\mbox{\tiny $+0.25-0.19$}}$ &  30  & 0.12 \\
MOO J0448-1705   & 0.960     & $4.58^{\mbox{\tiny $+0.34-0.34$}}$ &  27  & 0.11 \\
MACS J0717.5+3745   & 0.550     & $2.24^{\mbox{\tiny $+0.22-0.20$}}$ &  27  & 0.11 \\
2XMM J0830+5241   & 0.990     & $4.00^{\mbox{\tiny $+0.63-0.59$}}$ &  18  & 0.07 \\
RDCS J0910+5422   & 1.100     & $3.19^{\mbox{\tiny $+0.26-0.21$}}$ &  24  & 0.09 \\
MOO J1001+6619   & 1.530     & $2.12^{\mbox{\tiny $+0.56-1.19$}}$ &  27  & 0.11 \\
MOO J1014+0038   & 1.210     & $3.12^{\mbox{\tiny $+0.16-0.15$}}$ &  23  & 0.08 \\
Zwicky 3146   & 0.291     & $8.16^{\mbox{\tiny $+0.44-0.54$}}$ &  15  & 0.06 \\
MOO J1046+2757   & 1.160     & $2.00^{\mbox{\tiny $+0.21-0.23$}}$ &  36  & 0.16 \\
MOO J1052+0823   & 1.410     & $1.93^{\mbox{\tiny $+0.31-0.35$}}$ &  23  & 0.08 \\
RX J1053.7+5735   & 1.260     & $5.19^{\mbox{\tiny $+0.21-0.19$}}$ &  18  & 0.07 \\
MOO J1054+0505   & 1.450     & $1.34^{\mbox{\tiny $+0.33-0.34$}}$ &  40  & 0.14 \\
MOO J1059+5454   & 1.190     & $2.58^{\mbox{\tiny $+0.06-0.06$}}$ &  27  & 0.11 \\
MOO J1108+3242   & 1.020     & $2.31^{\mbox{\tiny $+0.19-0.20$}}$ &  21  & 0.09 \\
MOO J1110+6838   & 0.900     & $2.02^{\mbox{\tiny $+0.16-0.16$}}$ &  34  & 0.12 \\
MOO J1142+1527   & 1.100     & $3.52^{\mbox{\tiny $+0.19-0.19$}}$ &  44  & 0.17 \\
MACS J1149.5+2223   & 0.540     & $8.12^{\mbox{\tiny $+0.30-0.30$}}$ &  37  & 0.15 \\
MOO J1322-0228   & 0.820     & $3.07^{\mbox{\tiny $+0.41-0.53$}}$ &  28  & 0.11 \\
MOO J1329+5647   & 1.430     & $3.56^{\mbox{\tiny $+0.20-0.20$}}$ &  46  & 0.15 \\
RX J1347.5-1145   & 0.451     & $7.03^{\mbox{\tiny $+0.45-0.45$}}$ &  48  & 0.19 \\
MOO J1354+1329   & 1.480     & $2.46^{\mbox{\tiny $+0.25-0.30$}}$ &  44  & 0.15 \\
MOO J1506+5136   & 1.090     & $3.09^{\mbox{\tiny $+0.29-0.29$}}$ &  36  & 0.14 \\
Abell 2052   & 0.030     & $7.37^{\mbox{\tiny $+0.70-0.70$}}$ &  65  & 0.26 \\
MOO J1554-0447   & 1.050     & $5.36^{\mbox{\tiny $+0.73-0.85$}}$ &  50  & 0.20 \\

\hline 
\end{tabular}
\end{table}

\begin{figure*}
    \centering
    \includegraphics[trim=0in 2.4in 0in 0in,width=\textwidth]{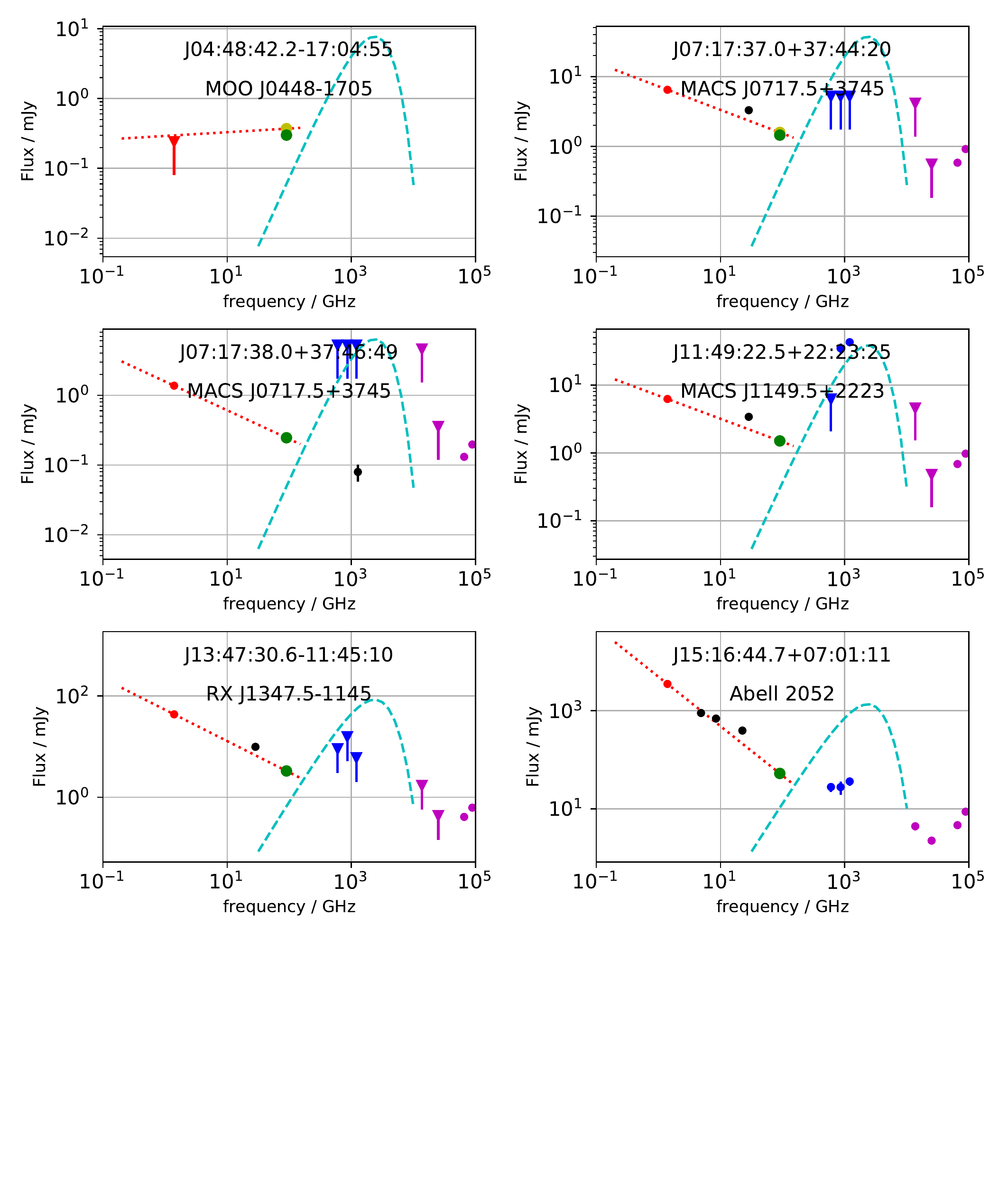}
    \caption{\label{fig:SEDs} Selected SEDs for the sources observed.  The MUSTANG2 flux densities assuming a point or an extended source at 10 arcsec resolution are shown by the green and yellow circles respectively -- only J04:48:42.2-17:04:55 shows strong evidence for extended emission so in the other plots the green points obscure the yellow ones. All points have error bars but many are too small to be easily seen.  The 1.4~GHz data from FIRST/NVSS are in red, where available Herschel/SPIRE data is in blue, black points are from the VLA and BIMA, and the four high frequency points (in purple) are from WISE - note the lower two frequencies of these four points are mostly upper limits (all upper limits are shown as triangles).  The fitted spectral index between 90~GHz and 1.4~GHz is the red dashed line.  A 40~K black body spectrum with a 90~GHz flux density set at 20 per cent  of the measured MUSTANG2 flux density is shown in cyan -- with the exception of J04:48:42.2-17:04:55 and J11:49:22.5+22:23:25 this emission is ruled out.  SEDs for other sources along with further notes are available as supplementary material.}
\end{figure*}    

\subsection{Source counterparts}
To place constraints on the spectral indices of these sources, searches for counterparts in radio and infrared surveys were carried out. A search radius of 9~arcsec was used except for lower resolution data sets, in which case half the beam width of the survey was used.  Where coverage was available, data at 1.4~GHz was obtained from the Faint Images of the Radio Sky at Twenty-cm (FIRST) point source catalog \citep{FIRST} while for all other clusters we used the NRAO VLA Sky Survey \citep[NVSS][]{Condon1998}. To account for the higher resolution of these surveys, integrated fluxes for any counterparts found were used. At the limits of the survey depths (1.0~mJy for FIRST and 2.4~mJy for NVSS) robust radio counterparts were found for 80 per cent  of the sources seen by MUSTANG2. In the case of Zwicky 3146 two of the sources are extended in the MUSTANG2 images but are marked as separate sources in the FIRST catalog.  In this paper we chose to combine the integrated FIRST fluxes. Infrared (3.4, 4.6, 12, \& 22~$\mu$m) counterparts for most sources were also found in the WISE survey \citep{WISE}; however, the density of sources in WISE is much higher than FIRST. Searches at locations randomly offset by 20~arcsec  from each of our sources had a match in WISE in over 50 per cent  of cases so it is likely that a significant number of WISE matches with MUSTANG2 sources could be chance alignments. Nevertheless, the analysis in this paper only uses the WISE sources to provide upper limits on emission from hot dust. Upper limits from counterparts that are chance alignments will be weaker upper limits than would be obtained without the chance alignment but in no case would these weaker limits change our conclusions. Few of the clusters in this paper were in the Herschel/SPIRE public archive, but where they were, counterparts or upper limits were obtained at wavelengths of 250, 350, and 500~$\mu$m.  A few of the point sources also had 28.5~GHz observations by BIMA \citep{BIMA} or VLA counterparts at 74~MHz \citep{Cohen2007}, 4.9, and 8.5~GHz \citep{Lin2009}. A summary of all these data can be found in Table~\ref{tab:src} while spectral energy density (SED) plots can be seen in Fig.~\ref{fig:SEDs} (selected sources only) and its extended version containing all the sources in the supplementary material.

These SED plots place limits on which emission mechanisms dominate at 90~GHz.  Any significant contribution from a hot ($\gg$40~K) thermal component is ruled out by the WISE data -- extrapolating the WISE flux densities to 90~GHz with any reasonable dust spectral index ($\beta>1)$ gives predicted emission an order of magnitude or more below the flux densities measured by MUSTANG2.  Conversely, the majority of the sources have a counterpart at 1.4~GHz and some of the better studied clusters  (e.g. Abell 2052 and RX~J1347.5-1145 in Fig.~\ref{fig:SEDs}) have additional radio data between 1.4 and 90~GHz. These data are mostly consistent with  spectral indices between $-0.1$ and $-1$ implying there is likely to be a synchrotron component in most of the sources measured by MUSTANG2.  In addition, Herschel data in some clusters such as MACS J0717.5+3745 put strong upper limits on a cold ($<$40~K) dust component, showing it contributes  20 per cent  or less of the flux density at 90~GHz.  
Many more of the sources are like those in the cluster MOO J0448-1705 on the top of Fig.~\ref{fig:SEDs} where  there are virtually no constraints on a cold dust component and a small change in the radio spectral index between 1.4 and 90~GHz would change the dominant emission mechanism in the MUSTANG2 data.  Five sources observed by MUSTANG2 show an inverted radio spectrum between 1.4 and 90~GHz but these SEDs could also be explained by the presence of a cold dust component or source variability.

\section{Extracting masses from millimeter wave surveys}\label{sec:ACT}
To see how point sources could affect masses recovered from tSZE surveys, a brief outline of the data analysis steps used by these surveys is needed.
Full details of the data analysis pipelines can be found in the relevant papers (e.g. \cite{Hilton2021,SPT_SZE,Planck_SZE}).  In this paper we concentrate on how masses are obtained from raw maps in the  recent ACT DR5 data release \citep{Hilton2021} which follows the multi-frequency matched filter approach in \citet{Melin2006}.  However, the methods used by other experiments and data releases are broadly similar. 

First, the magnitude of the tSZE (in units of flux density) varies with frequency as given by :
\begin{equation} \label{equ:g(x)}
    g(x) = \frac{x^4e^x}{(e^x -1)^2} \left(x \frac{e^x+1}{e^x-1} - 4\right)[1 + \delta(x,T_e) ]
\end{equation}
where $x$ is the dimensionless frequency defined as $ h \nu / k_\text{\tiny B} T_\text{\tiny CMB}$, $\delta(x,T_e)$ is a relativistic correction which is at most a few per cent and, in the analysis presented in this paper, can be ignored, and $T_e$ is the electron temperature.  Below $\sim$218~GHz, $g(x)$ is negative meaning that, at these wavelengths, clusters show up as decrements in the microwave background temperature (0.05--1~mK at 90~GHz).  Because of the unique spectral shape of the tSZE, it is possible to separate out the tSZE signal using multifrequency observations -- in the case of ACT DR5, 90 and 150~GHz are used to find clusters.  Other ground based experiments use similar bands while, due to the lack of atmospheric absorption, space based experiments such as {\it Planck} \citep{PlanckInstrument} can have wider and more complete frequency coverage.  Sets of matched filters (matched to $g(x)$ and the cluster profile) are used to extract maps of peak Compton-$y$ with different filter sets being used to obtain maximum signal-to-noise (SNR) on clusters in different mass and redshift ranges \citep{Hilton2021}.  
Because of how the magnitude of the tSZE changes with wavelength and the relative noise in the maps, frequencies around 90~GHz contribute most to the sensitivity of the ACT Compton-$y$ maps.  For most calculations \citet{Hilton2021} use a reference filter set optimized for clusters with $M_{500c}$ having an angular extent on the sky of 2.4 arcmin. The peak Compton-$y$ recovered from this filter set is referred to as \yc.  

With \yc\ calculated, the clusters are found by making cuts at fixed SNR ($4\sigma$ in the case of DR5).  However, because of  the log-normal nature of intrinsic scatter in the \ym\ relationships and the steepness of the cluster mass function, a simple inversion of the relationship is not used to evaluate mass directly.  
Instead, \citet{Hilton2021} find the most likely mass given the cluster's redshift and our knowledge of the intrinsic scatter.  For a given survey, the better our knowledge of the intrinsic scatter (which is potentially dependent on redshift) the more accurate the recovered masses will be. Any measurements of the causes of the intrinsic scatter in the measured $y$ values, such as the effects of point sources, apply across all tSZE cluster experiments current and future. 

\subsection{The effects of point sources}\label{sec:sims}
As the tSZE signal at the frequencies 
experiments such as ACT are most sensitive to is negative, central point sources will have the effect of `infilling' some of the tSZE signal. This will increase the scatter of the masses obtained by such surveys. Many authors \citep[e.g. ][]{Lin2007,SPT_src,Lin2009} have made calculations of the magnitude of this effect by equating source flux density to equivalent Compton-$y$. In most cases, data at tSZE frequencies are not available and extrapolations over two decades in frequency need to be made to predict the source population.  Small errors in these extrapolations can have a large effect and there is evidence that radio sources in the centers of clusters have different spectra indices than typical sources \citep{Coble2007}.  \citet{SPT_src} circumvent  this problem by looking for sources in their low resolution SPT maps  in the directions of X-ray clusters.  Cross correlations with the Sydney University Molonglo Sky Survey (SUMSS) at 843~MHz \citep{Murphy2007} were used to build a comprehensive model of the source populations within the virial radius of clusters. At z=0.25, this model predicts that 0.5 per cent of $3{\times}10^{14}\mbox{M}_\odot$\ clusters would be totally infilled at 150 GHz, rising to 1.5 percent at 90 GHz -- the inclusion of the higher frequency data giving a result 6 times lower than that of \citet{Lin2007}. However, these techniques which simply add positive cluster flux to the negative tSZE flux, do not fully take into account the matched filter in the cluster finding pipeline described above.

\begin{figure*}
    a\includegraphics[width=0.65\columnwidth]{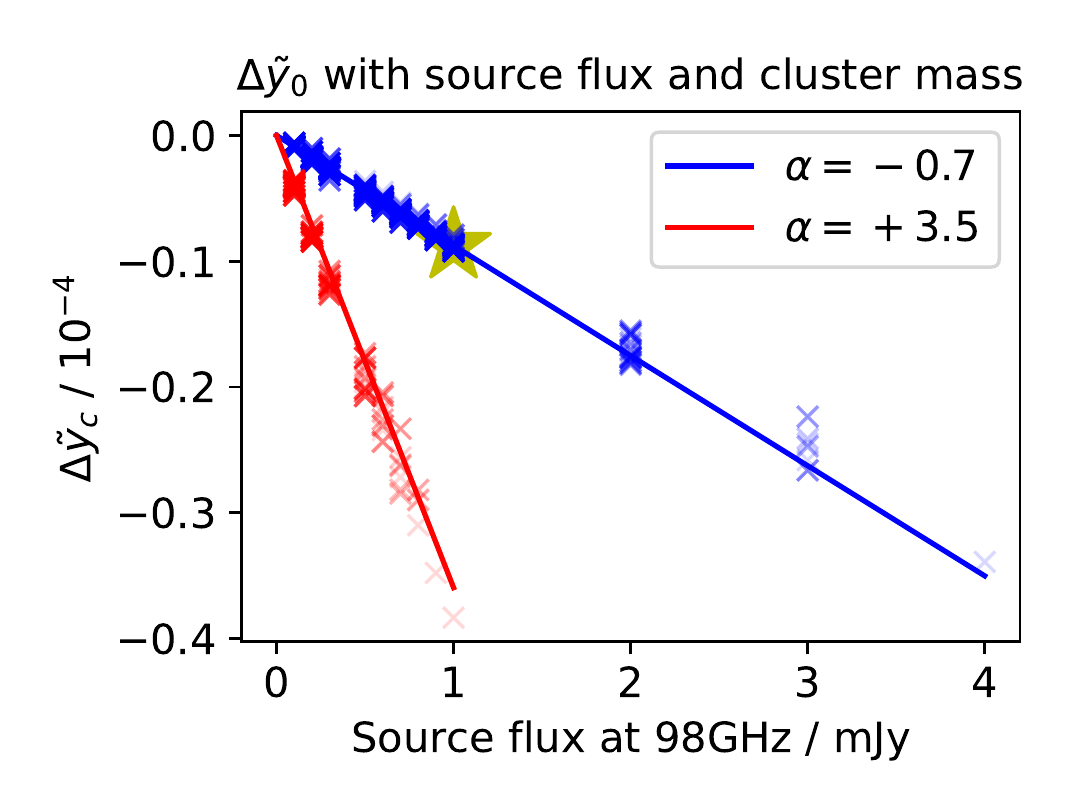}
	b\includegraphics[width=0.65\columnwidth]{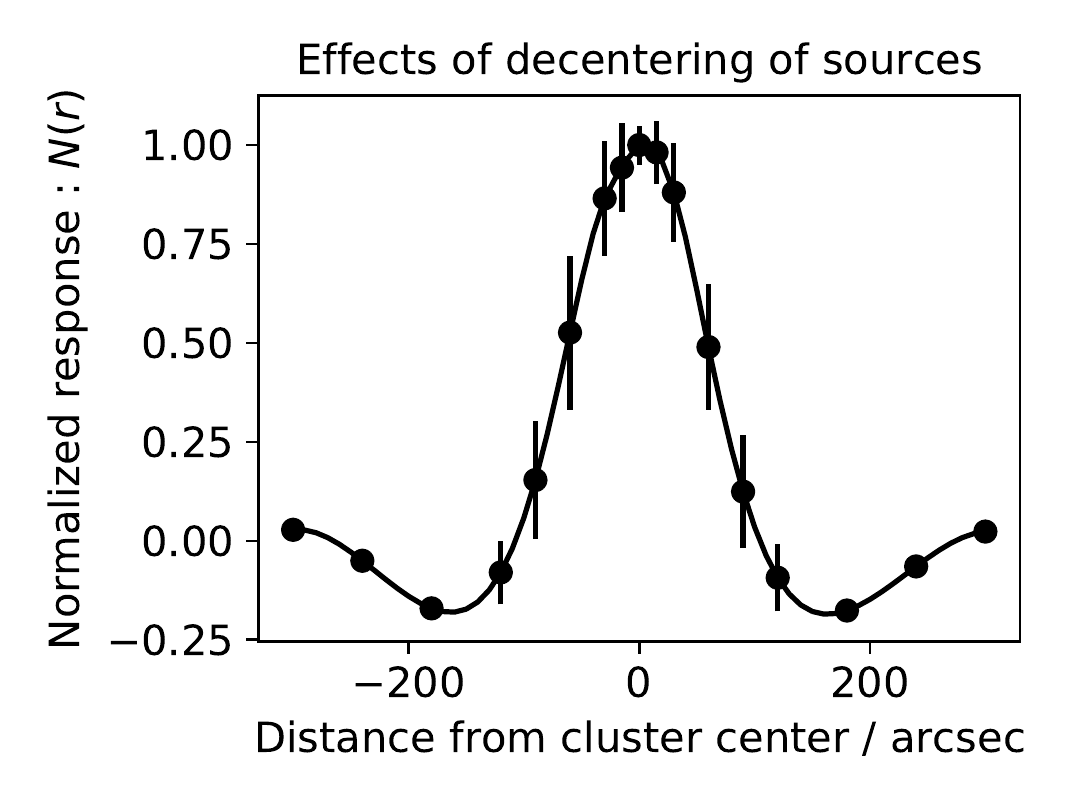}
	c\includegraphics[width=0.65\columnwidth]{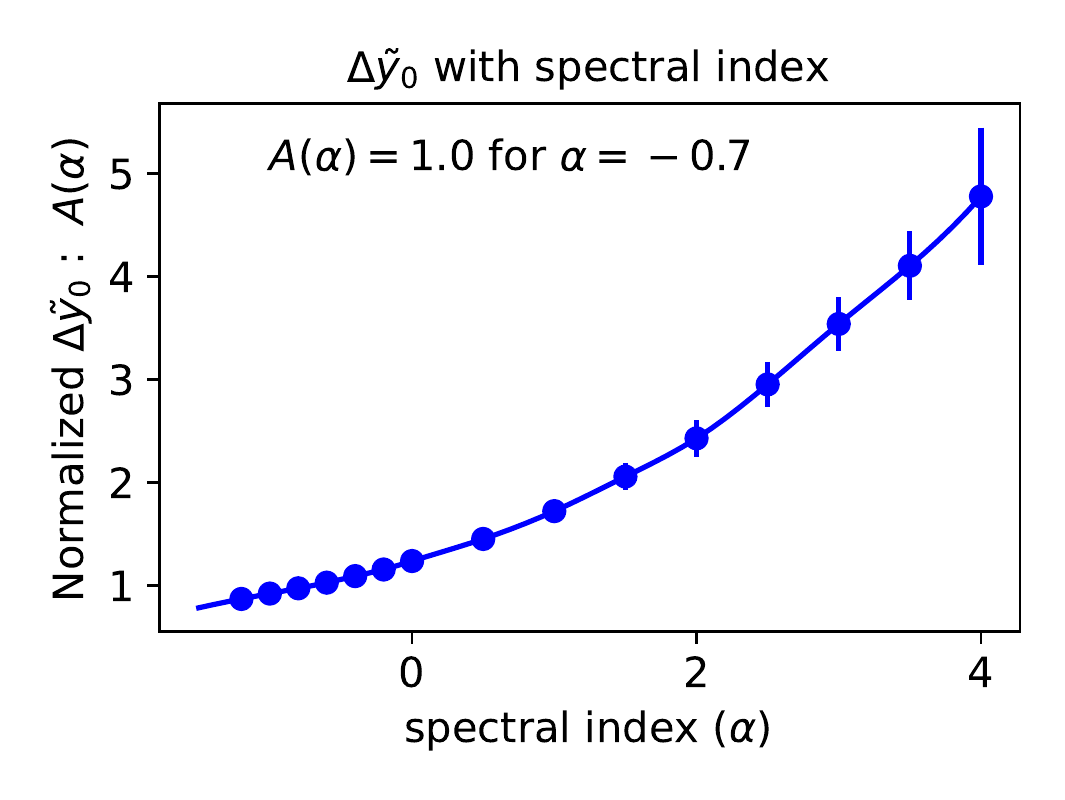}
	\caption{\label{fig:sim_results}Results from injecting sources to simulated clusters with different masses and redshifts. left: How \yc\ changes with the flux density of sources placed in the center of each cluster and the cluster mass (more massive, higher significance clusters are plotted in fainter colors). Sources with spectral indices of $-0.7$ and $+3.5$ are plotted in blue and red respectively. For a given spectral index the change in \yc\ is linear with source strength and independent of cluster mass. Simulated clusters were recovered from the maps with significances between $4.5\sigma$ (dark colours) to $10\sigma$ (lighter points).  The yellow star is for a 1~mJy, alpha=-0.7 source used to normalize the other plots; Center: How the average change in \yc\ varies as you move the reference source away from the center of the cluster normalized to a peak of 1 when the source is in the center. The results are symmetrical around r=0~arcsec; Right: The effects of spectral index normalized to a value of 1 for the reference source, averaged across all simulated clusters.}
\end{figure*}

To better quantify the effects of point sources on DR5's measurement of \yc, simulations were carried out; 25 simulated clusters with masses ranging from $M_{500c}{=}$2.1--$6.5{\times}10^{14}~\mbox{M}_\odot$ (corresponding to typical detections of 4.5--10$\sigma$) and redshifts between 0.145 and 1.85 (the range of redshifts in DR5) were placed at random locations (avoiding the positions of known clusters and sources) in a single $12.6^\circ{\times}7.3^\circ$  tile drawn from the 90 and 150 GHz ACT DR5 maps. Cluster profiles from \citet{Arnaud2010} were assumed and the maps were run through the ACT cluster detection pipeline \citep{Hilton2021}.  This process was then repeated with fake point sources added to the maps and the recovered properties of the simulated clusters with and without sources present where compared.   Initial tests placed sources with spectral indices of $\alpha=-0.7$ (a radio source) or $\alpha=3.5$ (a dust source) in the centers of the clusters. The results, shown in Fig.~\ref{fig:sim_results}a show that the change in \yc\ is independent of cluster mass and is linear with source flux density up until the source changes the recovered \yc\ by approximately 30 per cent. Sources that changed \yc\ by more than 30 per cent resulted in recovered \yc\ values with low SNR and high scatter.  If these were real clusters, most would not have made the SNR cut to be included in the DR5 catalog so these points were dropped from the simulations.  For the purposes of further analysis we adopt a reference source as having a spectral index of $\alpha=-0.7$ and a 90~GHz flux density of 1~mJy which gives a change in \yc\ of $-8.76{\times}10^{-6}$ (12 per cent for a typical $M_{500\text{c}}$=$2.5{\times}10^{14}~\mbox{M}_\odot$\ cluster). 

As not all sources will be in the center of clusters, more simulations were carried out adding the reference 1~mJy, $\alpha=-0.7$ source to the clusters at distances between 0 and 300~arcsec from their known centers.  After discarding data points where the source infill exceeded 30 per cent , the resulting $\Delta_{\tilde{y}_0}$ was found to be independent of cluster mass and an average could be taken across all simulated clusters to obtain the normalized response function $N(r)$ shown in Fig.~\ref{fig:sim_results}b. The independence of $N(r)$ from cluster mass is expected as it represents the compact source response of the matched filter used to calculate \yc\ not the filter's response to the cluster.  The size and shape of $N(r)$ falls between the 90~GHz and 150~GHz components of the matched filter.
 Past a radius of 59~arcsec, a source contributes less than half the change in \yc\ than it would in the center.  Also, past  104~arcsec, the shape of the matched filters used to find clusters means that a positive source will in fact add to the negative tSZE signal resulting in some scatter to higher masses.  At a radius of 220~arcsec the response to a source is still above 10 per cent  (but with the opposite sign) of that of the same source in the center of the cluster - even though this is far outside the $R_{500}$ of most clusters.  The above measurements from MUSTANG2 show there are many sources in this region. 

When calculating $N(r)$, note that $r$ refers to the angular distance of a source from the known center of the cluster not the measured position reported by the data analysis pipeline.  A strong source that significantly affects the measured \yc\ can shift the measured cluster location. For off-centered sources less than 104 arcsec from the true cluster center, the measured cluster location will move away from the source and calculations of $N(r)$ using the measured positions will be biased low.  However, for changes in \yc\ less than 20 per cent, the simulations in Sec.~\ref{sec:sims} show this effect is less important than variations in the measured locations of clusters due to map noise (${\sim}10$~arcsec for a cluster with SNR=8 -- see Figure~5 of \citet{Hilton2021}). Consequently, regardless of the presence of a source, the measured cluster locations can be used.  

From the first simulations, dusty sources had a much larger effect on \yc\ than radio sources of the same 90~GHz flux density. This is due to the higher flux density of the dusty source when extrapolated to the 150~GHz ACT band. To explore this in more detail, the spectral index, $\alpha$, of fake sources placed at the center of the clusters were changed between -1 and +4. The results of these simulations can be seen in Fig.~\ref{fig:sim_results}c. When normalized so that the reference source has an amplitude of 1, the results can be represented by the normalized function $A(\alpha)$.  Taking all this together, for any given source with a known distance from the cluster center $r$, flux density in the ACT 90~GHz band $I$ (in mJy), and spectral index $\alpha$, the difference in \yc\ can be written as:
\begin{equation} \label{equ:dy0}
\Delta_{\tilde{y}_0} = I \:\delta_{\tilde{y}_0}\: N(r)\: A(\alpha)
\end{equation}
where $\delta_{\tilde{y}_0}= -8.76{\times}10^{-6}$ is the reference value reported above for a 1~mJy source with a spectral index of $-0.7$. 

\section{Source Contamination in Clusters}\label{sec:results}
In this section we use Equation~\ref{equ:dy0} to predict the change in the measured \yc\ for different cluster samples.  To begin with, we use only the low frequency radio data to explore the expected contamination of the ACT DR5 sample under different assumptions.  We then use the flux densities measured by MUSTANG2 to calibrate this relationship.  After this we expand the analysis to cluster samples selected by other observing techniques.  

\subsection{Extrapolation from 1.4~GHz}\label{sec:1.4GHz}
\begin{figure}
    \centering
    \includegraphics[width=\columnwidth]{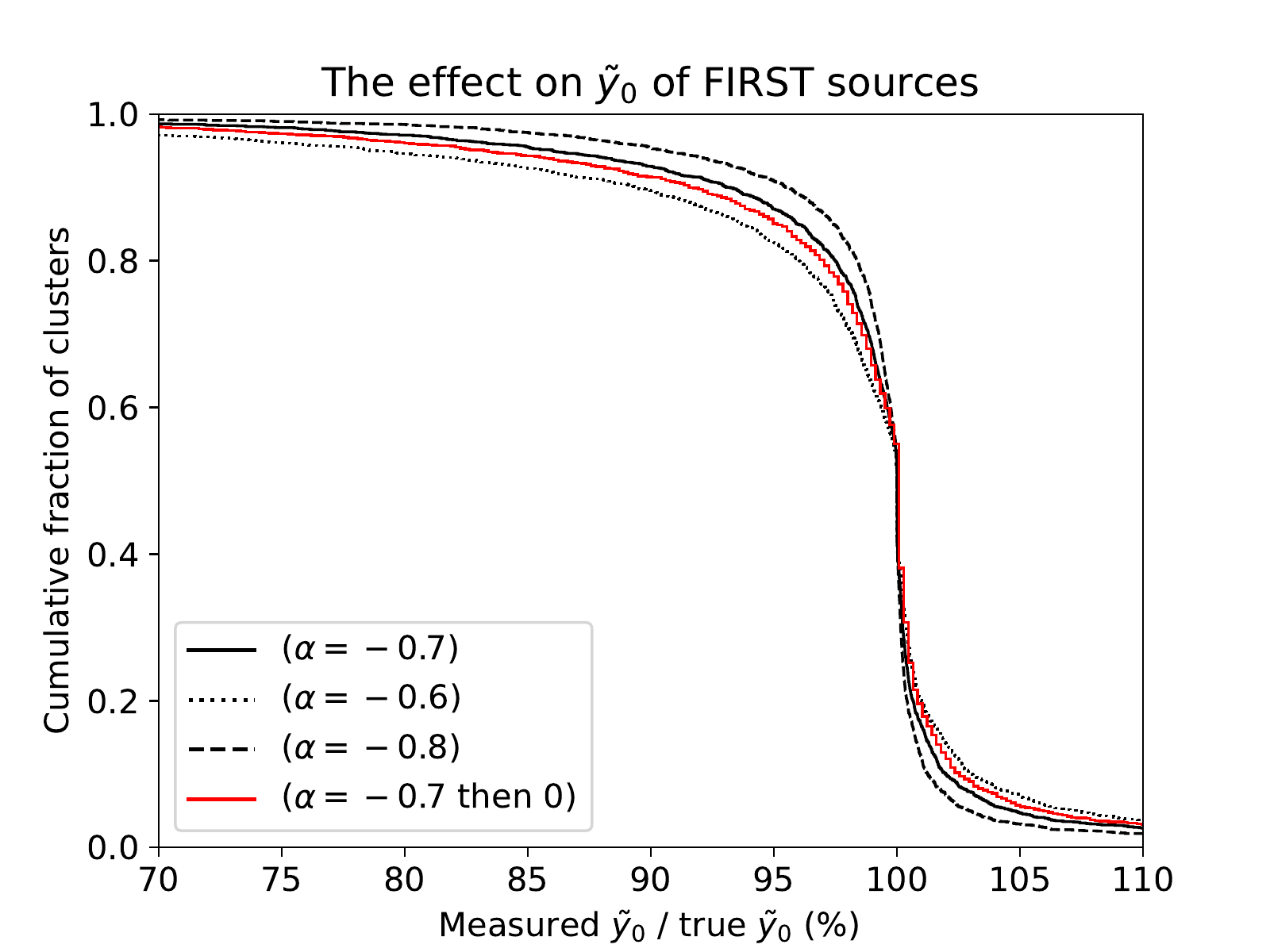}
    \caption{\label{fig:FIRST}Cumulative histograms of the percentage change in \yc\ from FIRST radio sources in DR5 clusters. The sources were extrapolated to the ACT 90~GHz band using different spectral indices (the black lines).  The red line uses a spectral index of $-0.7$ that flattens to 0.0 at 98~GHz.    
    }
\end{figure}

Using a search radius of 5~arcmin centered on each DR5 cluster in the FIRST survey footprint, we found 1.4~GHz flux densities for all sources above the FIRST detection threshold of 1~mJy.  In this subset of 2138 DR5 clusters, 1947 of them had one or more FIRST sources.  For these clusters, we made the large extrapolation from  1.4~GHz to the ACT 90~GHz band (actual central frequency 98~GHz) to find the flux density used in Equation~\ref{equ:dy0}. As there are no radio surveys at intermediate frequencies with sufficient resolution and sensitivity for the majority of the observed clusters, we simply assumed spectral indices of $-0.6$, $-0.7$, and $-0.8$ and assumed these spectral indices were constant across the ACT bands. The distance of each source from the cluster center was calculated and the fractional change in \yc\ found using Equation~\ref{equ:dy0}. Histograms of the results are shown in Fig.~\ref{fig:FIRST}. 

 For a spectral index of $-0.7$, we find that a significant number of clusters ($\sim$5 per cent ) have measured \yc\ 5 per cent  {\it above} their true values.  These are due to sources located more than 104~arcsec from the cluster centers and is an effect not predicted by calculations using simple aperture photometry. Although less important than the approximately equal number of sources found close to the cluster center this is a non-negligible effect.  

The reduction in the observed \yc, due to sources closer to the cluster center, has a long tail. When assuming a spectral index of $-0.7$, 2.8 per cent of clusters have a reduction in \yc\ of more than 20 per cent  which is broadly in line with predictions of \citet{Lin2007} who predict that this number is less than 3 per cent  of clusters. 
At lower contamination fractions, the number of clusters affected is much larger with 13.1 per cent  of clusters predicted to have a measured \yc\ reduced by 5 per cent.  However, the most important result is the sensitivity to spectral index of these numbers. Changes in the assumed spectral index of 0.1, far less then the typical scatter in spectral indices, can result in a factor of 2 change in the number of clusters affected, showing the importance of high frequency, high resolution point source searches within clusters in order to measure source flux densities directly. 

This sensitivity to spectral index of the predicted amplitude of these sources in the ACT 90~GHz band is driven by the large extrapolation from 1.4~GHz. Even when data at intermediate frequencies (5--20~GHz) are available, the flatter spectrum sources that are more likely to be bright at tSZE wavelengths are more likely to be variable \citep{ODea1998}.  As data at different frequencies can be taken years apart, accurate  extrapolations can be problematic.
Also, the spectral index $\alpha$ in Equation~\ref{equ:dy0} is the {\it local} spectral index between the frequency bands used to measure the tSZE.  In the case of ACT, these frequencies are in a range where many sources start to be dominated by dust, so their spectral index may change with frequency. Radio sources can also change spectral index as, at higher frequencies,  flatter-spectrum radio cores can start to dominate over the steep-spectrum radio lobes \citep{Whittam2017}. To test our sensitivity  to such spectral index changes,  the 1.4~GHz flux densities were extrapolated to the ACT 90~GHz band using a spectral index of $-0.7$ (to obtain $I$ in Equation~\ref{equ:dy0}) and above this frequency a flat spectrum of $\alpha=0$ was assumed (in $A(\alpha)$).  The results (the red line in Fig.~\ref{fig:FIRST}) are different from when a constant spectral index is assumed, demonstrating the need for additional data to constrain any dust contribution to the mm-wave spectrum of sources. 

\begin{figure}
    \centering
    \includegraphics[width=\columnwidth]{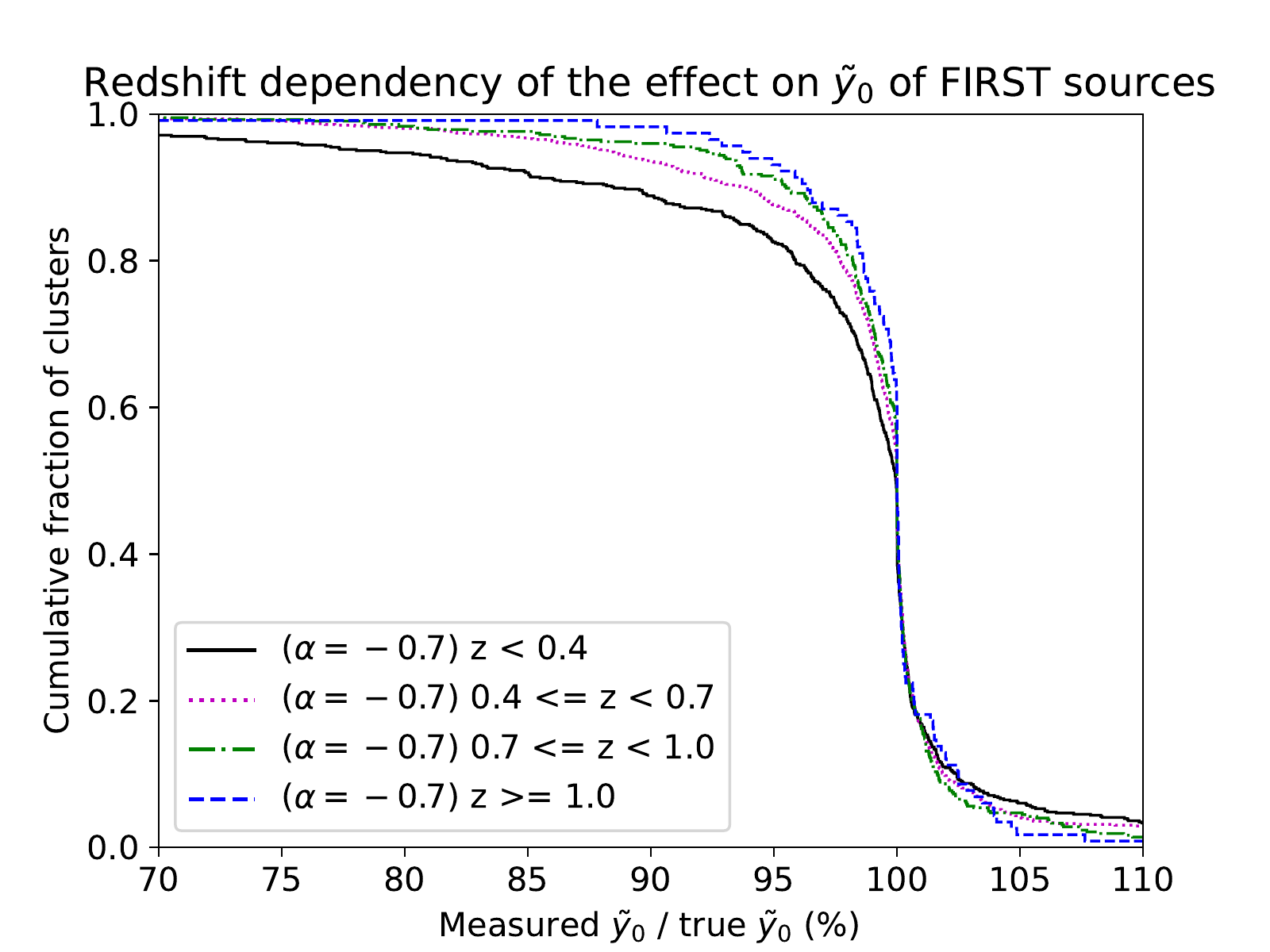}
    \caption{\label{fig:DR5redshift}Cumulative histograms of the percentage change in \yc\ predicted from 1.4~GHz source counts, broken down by cluster redshift.}
\end{figure}

As the number of DR5 clusters with FIRST coverage is large, it is possible to bin clusters by redshift. Fig.~\ref{fig:DR5redshift} shows an example for an assumed spectral index of $-0.7$.  As would be expected if the majority of these FIRST sources were associated with the clusters then, due to redshift dimming, clusters at redshifts below 0.4 have significantly more contamination from point sources. This trend extends all the way to redshifts past z=1 with redshift dimming more than making up for possible increases in source counts or luminosity at higher redshifts.

\subsection{MUSTANG2 constraints on ACT  $\Delta\tilde{y}_0$}
We return now to the MUSTANG2 results.  As stated in Section~\ref{sec:M2obs} many of the sources measured by MUSTANG2 are clearly dominated by radio emission so it makes sense to use the spectral index calculated between 1.4 and 90~GHz to make the small extrapolation from the center of MUSTANG2's band (90~GHz) to ACT's (98~GHz) as well as the value for $\alpha$ (the spectral index within the ACT bands).  In other sources, where a cold thermal component cannot be ruled out, a 1.4~GHz radio counterpart is often present and it is likely that the transition from being radio to dust dominated is close to MUSTANG2's measurement so $\alpha$ will be somewhere between a typical dust and a radio index. Any dust component of the emission at 90~GHz will have the effect of increasing the spectral index calculated between 1.4 and 90~GHz over that calculated where no dust contribution is present, giving a value between that of dust and radio.  For these sources, the spectral index calculated between 1.4 and 90~GHz is probably the best estimate for $\alpha$\ from the available data.  In the cases where no 1.4~GHz counterpart source was found, the point source sensitivity limit of the appropriate catalog was used instead.  Calculated values of the resulting spectral index are in Table~\ref{tab:src} and range between $-1.01$ and 0.24 with a median value of $-0.460$, comparable to the median value of $-0.66$ for sources in clusters found by \citet{Coble2007}.   

To find the value of \yc\ that ACT would have measured if it had observed the MUSTANG2 clusters (many of which are outside the ACT survey area), we use equation 5 from \citet{Hilton2021} :
\begin{equation}
    \tilde{y}_0 = 4.95{\times }10^{-5} E(z)^2 \left( \frac{{M}_{500c}}{3\times 10 ^{14}} \right)^{1.08} Q({M}_{500c},z) f_{\rm rel}
    \label{equ:y0}
\end{equation}
where ${M}_{500c}$ is the cluster mass, $E(z)$ is the evolution of the Hubble parameter with redshift (e.g. $\sqrt{\Omega_m(1+z)^3+\Omega_\Lambda}$), $Q({M}_{500c},z)$ is a function that describes the mismatch between the clusters angular size and the 2.4~arcmin matched filter used to calculate \yc. $Q({M}_{500c},z)$  becomes significant for large clusters at low redshifts. $f_{\rm rel}$ is a relativistic correction  which is far less than the assumed errors in ${M}_{500c}$ (typically 1--2 per cent) and so is taken to be 1. The cluster masses used are those in Table~\ref{tab:clusters} which are derived from non-parametric cluster profile fits to the MUSTANG2 data using the method described in \citet{Romero2020} and \citet{Dicker2020}.  The fits are carried out on the calibrated detector timestreams and the point sources are included in the fits.

\begin{figure}
    \centering
    \includegraphics[width=\columnwidth]{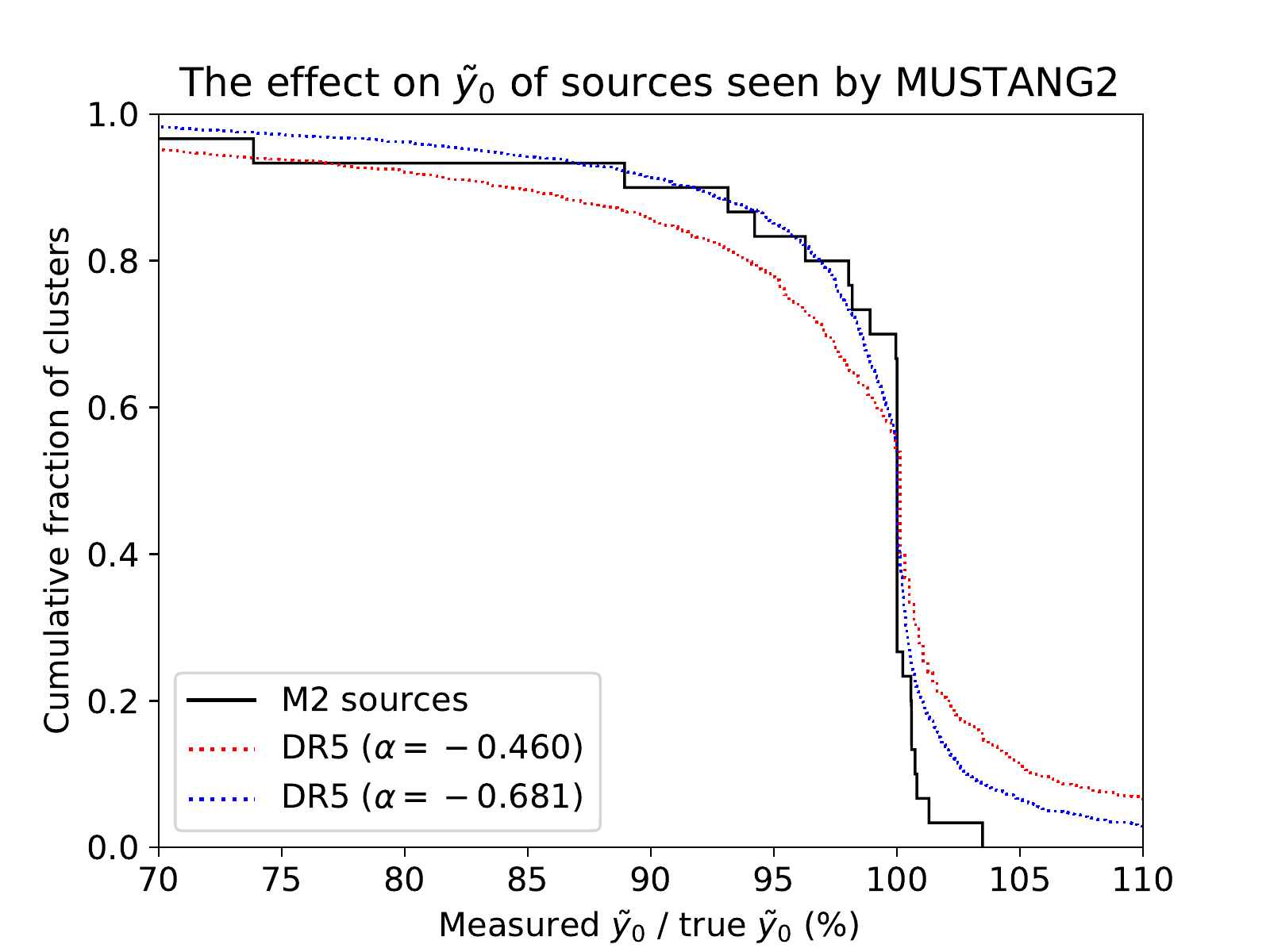}
    \caption{\label{fig:results}A cumulative histogram of the percentage change in \yc\ for the sources found by MUSTANG2. The dotted histograms are calculations of a sample of DR5 clusters matched in redshift to the MUSTANG2 sample with point source flux densities extrapolated from FIRST 1.4~GHz flux densities using spectral indices of -0.460 in red (to match the median spectral index of the MUSTANG2 sources) and $-0.68$  in blue (which provides a better match to the histogram measured by MUSTANG2). }
\end{figure}

The \yc\ values calculated from equation~\ref{equ:y0} and spectral indices from Table~\ref{tab:src} were used to calculate the fractional change in \yc\ shown in Fig.~\ref{fig:results}. 
Of our sample of 30 clusters, 5 have a change in \yc\ of more than 5 per cent (MACS J0717.5+3745=6 per cent; ACT-CL J0326-0043=7 per cent; RX J1347.5-1145=12 per cent; MOO J1554-0447=26 per cent; Abell 2052=395 per cent -- it appears as a point source in Fig.\ref{fig:maps}.)

To properly compare the predicted effects of point sources between the DR5 and MUSTANG2 samples, the dependency on redshift needs to be taken into account. There are enough DR5 clusters in the FIRST region to bin into redshift bins of width $\Delta z = 0.1$ while maintaining a meaningful sample (${\gg}10$) in each bin over the redshift range z=0.1 to 1.3.  Histograms were taken within each bin and then added together with weights matched to the redshift distribution of the MUSTANG2 clusters.  Fig.~\ref{fig:results} shows the histograms obtained by extrapolating the FIRST sources found in a redshift matched sample of DR5 clusters using the median spectral index found by MUSTANG2.  This predicts a larger effect on \yc\ for the DR5 sample than the measured values from MUSTANG2.  The best match (in a least squares sense) between the two samples uses a spectral index of $-0.68$.  A best fit spectral index steeper than the median value found in sources detected by MUSTANG2 reflects the large amount of scatter in spectral indices -- a significant number of the FIRST sources will have steep spectral indices and fall below the MUSTANG2 detection threshold effectively biasing the median value found by MUSTANG2 high.  
For the purpose of predicting the contamination of the DR5 sample, using a single spectral index across all redshifts, source locations within clusters, and cluster masses is clearly an approximation. A larger survey for sources in clusters over wider cluster redshift and mass ranges would allow us to test for effects such as source evolution and to build a model that takes them into account.   

\begin{figure}
    \centering
    \includegraphics[width=\columnwidth]{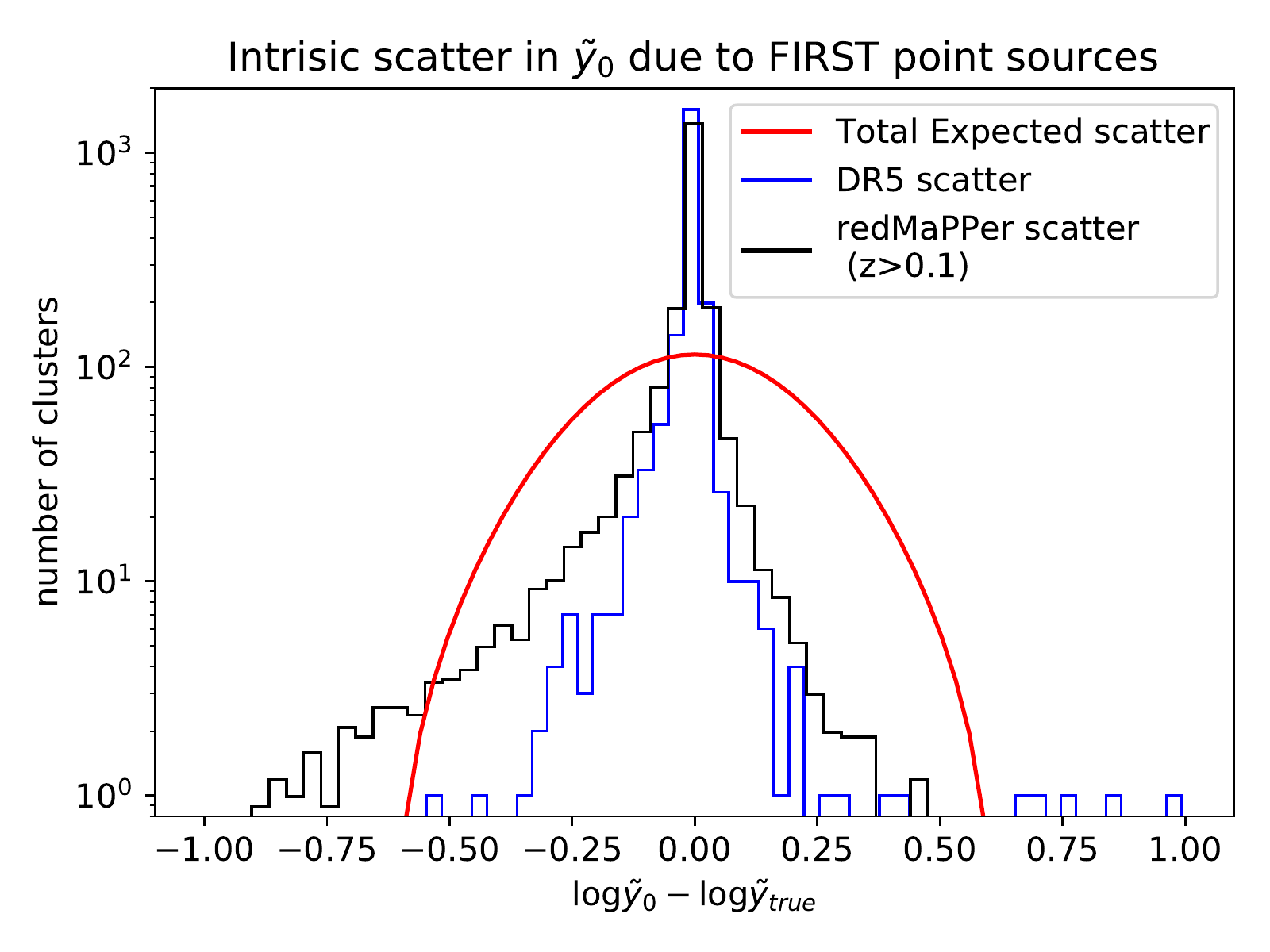}
        \caption{\label{fig:intrisic}Histograms of the predicted scatter in \yc\ due to FIRST point sources for redMaPPer (black) and DR5 (blue) clusters.  A single spectral index of $-0.68$ , the best fit for the MUSTANG2 data, has been used and the samples have been scaled to the same size.  For comparison, in red, is a fit to the total (20 per cent) intrinsic  scatter of the clusters that were used to calibrate the \ym\ relationship from figure~9 in \citet{Hasselfield2013} (normalized for sample size). The scatter due to point sources in DR5 is 6 per cent  compared to 11.3 per cent  for the redMaPPer clusters.  Although the scatter due to point sources is smaller than the total scatter there are long tails. The difference between the histograms on the negative side is evidence that a significant number of clusters are missing from the DR5 sample due to point sources.}
\end{figure}

However, taking this result at face value, contamination fractions similar to \citet{Lin2007} are obtained,  3 per cent  of clusters have more than a 20 per cent decrease in Compton-$y$.  Unlike \citet{Lin2007} we also predict another 3 per cent will have a 10 per cent increase. In addition, it is possible to calculate the intrinsic scatter in DR5 clusters due to point sources alone and compare it with the value of the scatter in the fitted \ym\ relationship of $\sigma (\log \tilde{y}_0) = 0.2$ \citep{Hasselfield2013}.  The results in Fig.~\ref{fig:intrisic} show that, while the peak in the intrinsic scatter due to point sources is much more narrow than that in \citet{Hasselfield2013}, there exist long tails with significant amounts of scatter. The scatter in the DR5 clusters due only to point sources is 6 per cent. However, tSZE surveys are inherently biased -- clusters with significant point source contamination will be missing from the DR5 sample making the scatter artificially low.  In the next section, comparisons with clusters selected by non-tSZE methods shows this effect is important.

\subsection{Comparisons between cluster surveys}\label{sec:comp}
In this section we examine how common point sources are between clusters selected by different survey techniques.  For comparison we choose the Meta-Catalog of X-ray detected
Clusters of galaxies \cite[MCXC][]{MCXC}, an X-ray catalog made by combining observations from many different ROSAT and {\it Einstein} cluster catalogs.  Although highly heterogeneous (in terms of depth and redshift ranges), this sample has 732 clusters in the FIRST region giving it better constraining power than some smaller but more pure samples. Like \citet{SPT_src}, we assign a generous 40 per cent  error to the masses of this survey.  To compare tSZE selected clusters to optically selected clusters we choose the SDSS DR8 redMaPPer catalog  which has over 22\,300 galaxy clusters in the FIRST footprint and a 21 per cent intrinsic scatter to true halo mass  when calibrated to {\it Planck} using the scaling relation from \citet{RedmapperIII}. The redMaPPer catalog is made using  an iterative algorithm that finds galaxy clusters using the red sequence \citep{RedmapperI}. There is significant overlap among the optical, X-ray, and ACT DR5 cluster samples, but the overlap is far from complete. The X-ray and optical catalogs extend to lower mass values but they are less sensitive than ACT at higher redshifts due to their approximately flux-limited natures.  DR5 clusters in the FIRST region have redshifts between 0.035 and 1.91 with a median value of 0.518, MCXC redshifts with FIRST coverage range from 0.0031 to 1.261 with a median value of 0.161, and redMaPPer clusters with FIRST coverage have redshifts between 0.062 and 0.94 with a median value of 0.368.  Direct comparisons between the MCXC / redMaPPer catalogs and the MUSTANG2 sample was not possible due to limited overlap in redshift.

To ensure the clusters were on the same mass scale, the MCXC and redMaPPer catalogs were searched to find co-detections with DR5.  Possible matches were identified as being less then 5 arcmin apart on the sky and within 0.1 in redshift. When more than one match was possible both pairs were rejected.  Over the region of the sky with FIRST coverage, we identified 100 DR5/MCXC and 983 DR5/redMaPPer potential matches.  The ratio of the DR5 mass to the MCXC/redMaPPer masses were calculated and the median and standard deviations of these ratios found. The MCXC clusters had a median mass 5 per cent higher than the DR5 clusters with a scatter of 43 per cent.  This scatter is consistent with our assumed value for the intrinsic scatter of the MCXC sample. The redMaPPer mass scale was 37 per cent higher than the DR5 clusters with a scatter of 39 per cent which, given the 21 per cent scatter in the redMaPPer mass richnesss relation \citep{RedmapperIII}, is higher than expected.  Similar mass discrepancies between optical and tSZE measurements of clusters have recently been pointed out by \citet{JackOS2021} and \citet{Myles2021} showed that, at low redshift, redMaPPer clusters have a richness dependent bias due to projection effects.  When averaged over richnesses, this is large enough to explain the additional scatter in the measured masses between DR5 and RedMaPPer co-detections as well as the higher median mass.  This is not something that can be corrected on a cluster by cluster basis, so  
 for an initial comparison of the effects of  point sources on clusters selected via different techniques, we simply scale the cluster masses in the MCXC and redMaPPer samples so that co-detections are on the same average mass scale.

\begin{figure}
    \centering
    \includegraphics[width=\columnwidth]{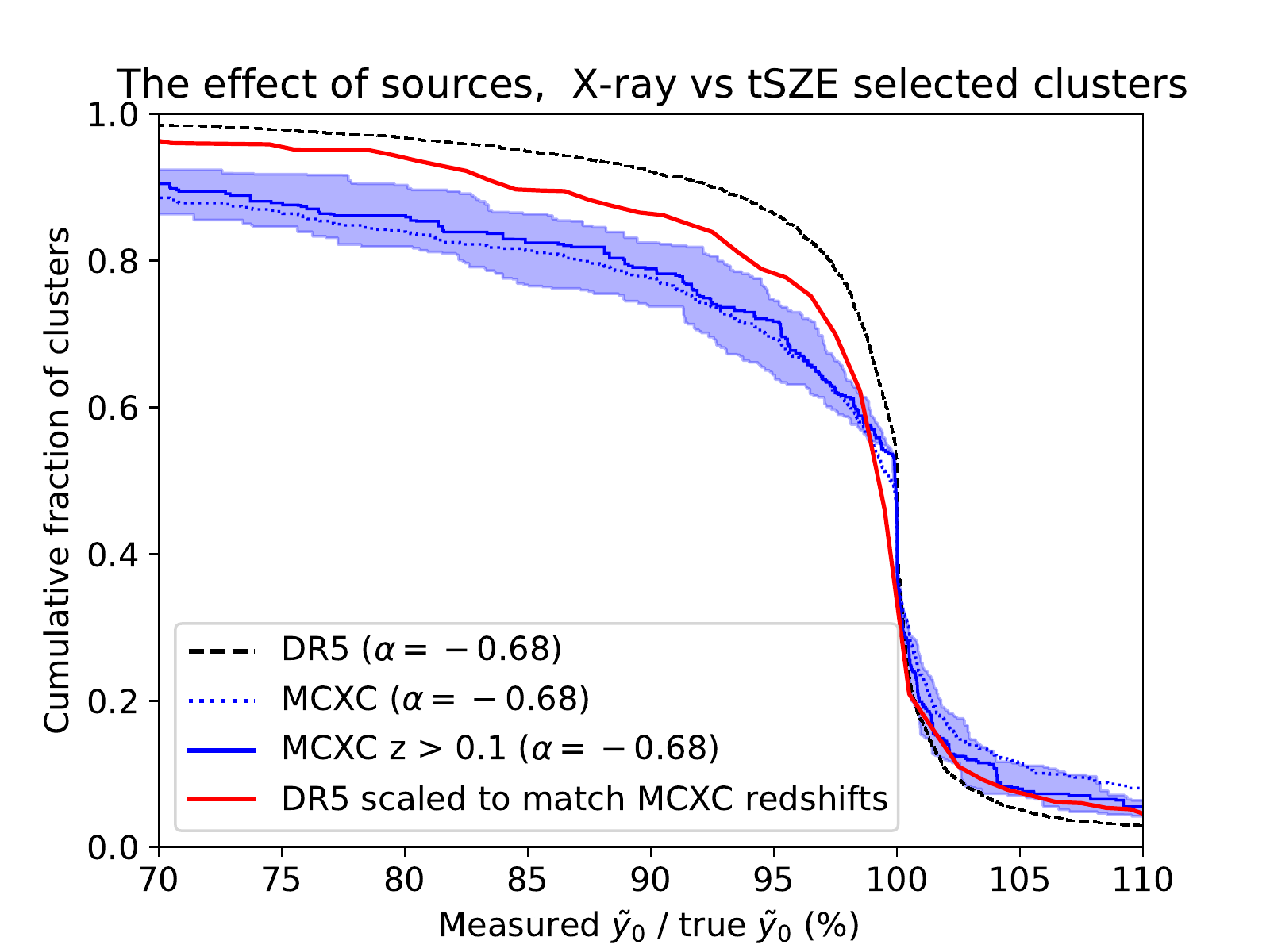}
    \includegraphics[width=\columnwidth]{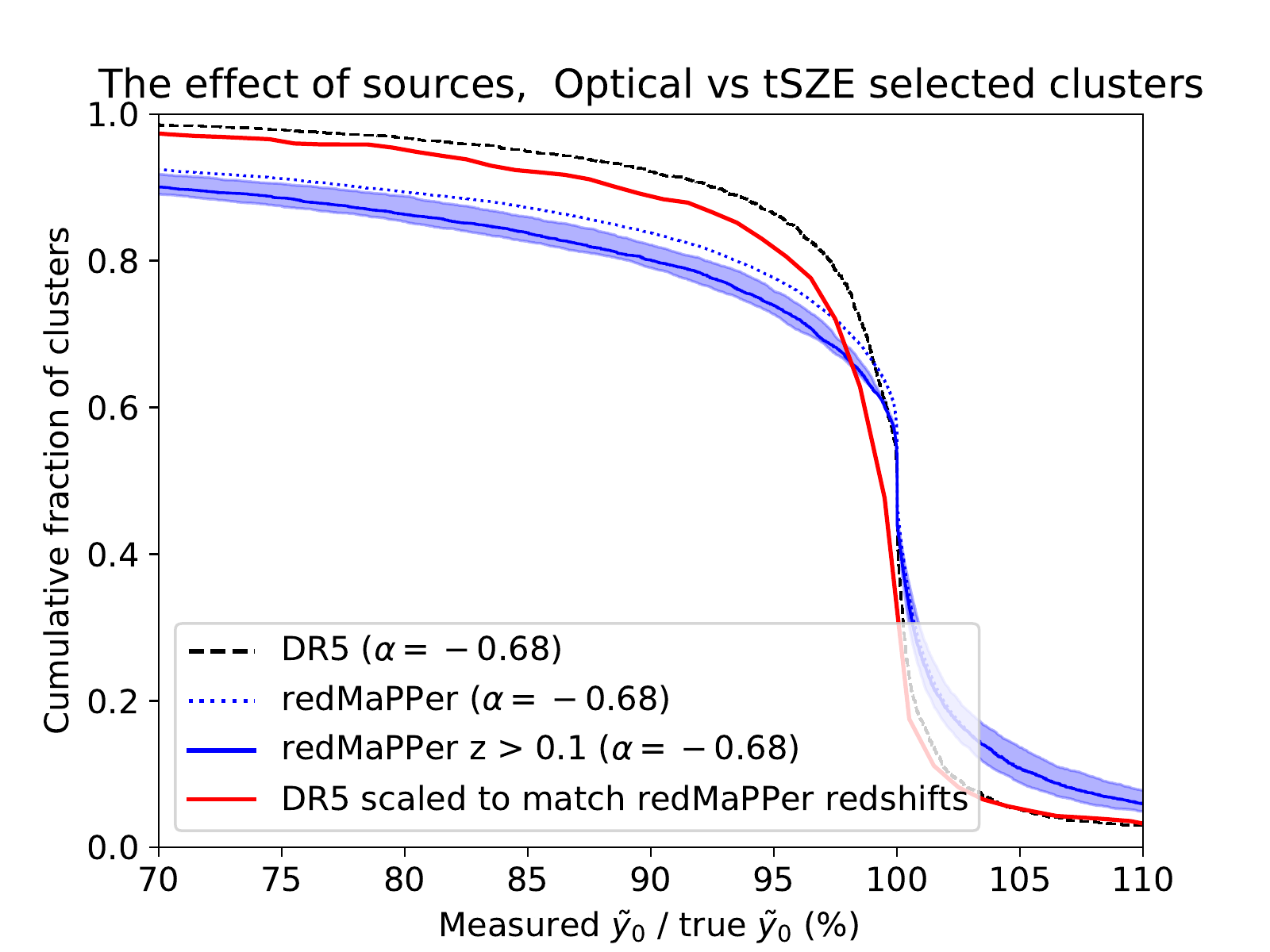}
    \caption{\label{fig:otherSamples} Predictions of the change in \yc\ due to point sources for clusters selected by other observational techniques. The dotted blue lines represent the complete samples while the solid blue lines include a cut to only include clusters above a redshift of 0.1, weighted to match the mass distribution of the DR5 catalog.  The shaded blue areas represent the uncertainties in the expected true \yc\ values which are based on the reported cluster masses. For comparison the entire DR5 prediction is plotted (the dashed black line) and a version of the DR5 catalog scaled to match the redshift distribution of the relevant survey is shown in red.  The top plot shows clusters taken from the MCXC X-ray survey while the bottom plot uses clusters from the redMaPPer catalog.}
\end{figure}

As with the DR5 sample, the FIRST point source catalog was searched for any sources located within 5~arcmin of the cluster centers and for each cluster $\Delta_{\tilde{y}_0} $ was calculated using Equation~\ref{equ:dy0}. Values for \yc\ were calculated using Equation~\ref{equ:y0} and the scaled cluster mass from the relevant survey.   Upper and lower limits on \yc\ were calculated for each cluster assuming the 40 per cent  and 21 per cent  errors in the masses for the MCXC and redMaPPer surveys.  Fractional differences that FIRST sources would have made to these \yc\ values assuming the  spectral index that was found to best match the DR clusters ($-0.68$) are shown in Fig.~\ref{fig:otherSamples}. For comparison, similar histograms of all DR5 clusters with FIRST data and a subset of these clusters chosen to match the redshift distributions of the other catalogs are also plotted. Weights have been applied so that the MCXC and redMaPPer surveys have a similar distribution of \yc\ values to DR5. 

Even allowing for errors in the cluster masses and cutting all clusters below z=0.1 (where ACT is less sensitive) there is significantly more point source contamination in the non-tSZE selected samples -- 13.5 per cent  of redMaPPer clusters have more than 20 per cent  contamination compared to just 4.9 per cent  of the redshift adjusted DR5 sample. For the MCXC sample, the difference is similar with 14.5 per cent  of the X-ray clusters having more than 20 per cent  contamination compared to only 6 per cent  for a DR5 sample adjusted to match the MCXC redshift distribution.  This can be explained by the fact that, on average, radio sources decrease the amplitude of the tSZE so some clusters with strong radio contamination will be scattered out of a tSZE sample.  Evidence of this happening can be seen in the intrinsic scatter of the redMaPPer clusters shown in Fig.~\ref{fig:intrisic}. Although the scatter in the DR5 and redMaPPer smaples are similar at positive values (where sources over 104 arcsec from the cluster centers increase \yc ), there are far more redMaPPer selected clusters on the negative side (due to sources close to the cluster center cancelling out the calculated tSZE signal).  This larger negative tail is expected from equation~\ref{equ:dy0} and the distribution of sources observed by MUSTANG2 (Fig.~\ref{fig:dist}).  Quantitative comparisons of the scatter in each sample are limited by calibrations and systematics such as uncorrected differences in the mass and redshift distributions. However, by taking the difference between the two histograms and assuming that the scatter in \yc\ due to point sources of the underlying cluster population is better described by the redMaPPer sample, it can be calculated that approximately 5~per cent of DR5 clusters could be missing due to point source contamination.  This is on this higher side of previous studies, such as \citet{SPT_src}, which have put this number between 1.8 and 5.6 per cent.  An observational program to compare the prevalence of sources at tSZE wavelengths in tSZE and non-tSZE selected clusters would better constrain this number.

\section{Conclusions}\label{sec:conclusions}
In this paper we have presented flux density measurements at the frequencies used in tSZE surveys of a population of radio sources that could cause reductions or enhancements in the measured \yc\ of clusters by a significant amount (5 per cent  and larger).  For massive clusters with high SNR, reductions result in a lower inferred mass but for smaller clusters this can result in non-detections. By comparing with optical surveys there are indications that undetected sources could be masking  5 per cent of clusters in the DR5 survey. 
Enhancements due to sources, caused by the shape of the matched filters used to find clusters, increase the intrinsic scatter in the \ym\ relationship -- sources almost 4~arcmin from the cluster centers can still have a 10 per cent  effect on the measured \yc .   Although these results have not been tested on other tSZE surveys, as similar data processing steps are used, the results are likely to be similar, warranting further investigation.

Because of the wide variation in spectral indices, using low-frequency radio surveys such as FIRST and NVSS, to remove these sources at the much higher frequencies of 90--150~GHz is very inaccurate.  As is shown in Fig.~\ref{fig:FIRST}, a spectral index change as small as 0.1 can double the predicted effect of a source on the central Compton-$y$ of a galaxy cluster.  The variation in spectral indices in the radio, seen in this work and others \citep[e.g.][]{PlanckSources}, is an order of magnitude larger than this.  Higher frequency measurements at tSZE wavelengths, such as those from MUSTANG2 presented in this paper, can greatly improve matters as there is no need to extrapolate source flux densities over two decades in frequency. However, the spectral index (and hence the dominant emission mechanism of the source) at the wavelengths used by cluster surveys will matter. We were able to find radio (1.4--28.5~GHz) counterparts for 80 per cent  of the sources detected by MUSTANG2 and these indicated that at frequencies around 90~GHz, the source population is dominated by radio sources and the spectral indices could be estimated using the 1.4~GHz data.  It is worth noting that, due to emission from cold dust,  at 90~GHz, the spectral index of some sources will vary with frequency. For these sources, calculations made using spectral indices calculated at radio wavelengths will have small but still significant change in the predicted difference in the central Compton-$y$.  Due to the small number of sources in the sample in this paper with multiple radio/submillimeter measurements it is not possible to make firm predictions on how common this effect is.  

A larger survey of several hundred clusters at the frequencies used by tSZE surveys would be valuable.  Resolutions better than 15~arcsec and, for the ACT DR5 data release, a depth of at least 0.7~mJy$\,$beam$^{-1}$ would enable all sources that could significantly bias \yc\ to be found. A shallower survey would still be useful as, with a large enough sample, it should be possible to extrapolate source counts. Such a survey should go out to at least 4~arcmin from the center of each cluster in order to find all sources of importance (those sources where $N(r)$ is non-negligible).  This is significantly larger than the field-of-view of instruments such as ALMA.  While a single frequency (e.g. 90~GHz) would be useful, follow up observations at frequencies such as 30 or 150~GHz of the sources found would make such a survey even more valuable as spectral indices within the tSZE frequency bands  could be calculated. These follow-up observations could be highly targeted and would not require large maps.  With a large enough sample, the distribution of spectral indices would be robust against issues such as source variability between observations.  It is also worth noting that the sample presented in this paper is dominated by clusters with redshifts greater than z=0.4.  The number of sources of different types evolves with redshift, for example, AGN in clusters are more common above z=0.4 \citep{Martini2009}. However, due to cosmological dimming, sources at higher redshifts will be fainter and affect the central Compton-$y$ by less than the increase in source counts would indicate. By splitting up the DR5 survey into redshift bins we showed that lower redshift clusters are, on average, more affected by sources.  A large survey should include clusters across all redshifts so it can better quantify this. 

The analysis in this paper does not take into account that, assuming the same rest-frame SED, redshifting will disfavor detection of synchrotron sources and favor detection of dusty sources at tSZE frequencies. With a larger survey for sources than in this paper it would be possible to look for how the source population changes with cluster mass and redshift and build a more complex model that could be used to make better use of current and future tSZE surveys.  Including clusters identified by different methods (e.g. optical, X-ray, and tSZE) would also help measure any biases caused by the different selection effects of each technique. Hints of some of the differences can be seen in the different amounts of point source contamination found when comparing MCXC and redMaPPer clusters with the DR5 sample. Comparisons of the source populations within clusters selected by different methods would also help quantify the number of clusters missed by tSZE surveys due to contamination. Better knowledge of the statistics of the point source population would feed back into the process of translating measured Compton-$y$ to cluster mass, not just for specific clusters in the ACT survey discussed in this paper but for other experiments underway and in the future.

The simulations presented in Section~\ref{sec:ACT} used only 90~GHz and 150~GHz data.  Data currently being taken by ACT includes data at 30 and 40~GHz and in upcoming data releases the 220~GHz data will have lower noise too. Future experiments such as the Simons Observatory will also have lower and higher frequency information. This opens up the possibility of using our knowledge of the point source population's typical spectra, number counts, distance from the cluster centers, and evolution with redshift to better detect (and possibly correct for) clusters with significant point source contamination by looking for discrepancies in the measured Compton-$y$ between frequency channels. Simulations similar to those presented in Section~\ref{sec:sims} of this paper would be an important part of this analysis.  Due to the distribution of spectral indices of sources and possible source variability with time, the identification of all clusters in a tSZE survey with high levels of contamination is not possible using low frequency surveys such as FIRST.

\section*{Acknowledgements}
MUSTANG2 is supported by the NSF award number 1615604 and by the Mt.\ Cuba Astronomical Foundation. 
This material is based upon work supported by the Green Bank Observatory. 
GBT data were acquired under the project IDs AGBT17A\_340, AGBT17A\_358, AGBT17B\_101, AGBT17B\_266, AGBT17B\_334, AGBT18B\_215, AGBT19B\_200, and AGBT20A\_290.
The Green Bank Observatory is a facility of the National Science Foundation operated under cooperative agreement by Associated Universities, Inc. 

The ACT project is  supported  by  the  U.S.  National  Science Foundation  through  awards  AST-0408698,  AST-0965625,  and  AST-1440226,  as well as awards PHY-0355328,  PHY-0855887 and PHY-1214379.
Funding was also provided by Princeton University, the  University  of  Pennsylvania, and  a  Canada Foundation for Innovation (CFI) award to UBC.
ACT operates in the Parque Astron\'{o}mico Atacama in northern Chile under the auspices of the La Agencia Nacional de Investigaci\'{o}n y Desarrollo (ANID; formerly Comisi\'{o}n Nacional de Investigaci\'{o}n Cient\'{i}fica y Tecnol\'{o}gica de Chile, or CONICYT).
The development of multichroic detectors and lenses was supported by NASA grants NNX13AE56G and NNX14AB58G.
Detector research at NIST was supported by the NIST Innovations in Measurement Science program. Computations were performed on Cori at NERSC as part of the CMB Community allocation, on the Niagara supercomputer at the SciNet HPC Consortium, and on Feynman and Tiger at Princeton Research Computing, and on the hippo cluster at the University of KwaZulu-Natal. SciNet is funded by the CFI under the auspices of Compute Canada, the Government of Ontario, the Ontario Research Fund--Research Excellence, and the University of Toronto.
Colleagues at AstroNorte and RadioSky provide logistical support and keep operations in Chile running smoothly. We also thank the Mishrahi Fund and the Wilkinson Fund for their generous support of the project.

JPH acknowledges funding for SZ cluster studies from NSF grant number AST-1615657, NS acknowledges support from NSF grant number AST-1907657, and ADH acknowledges support from the Sutton Family Chair in Science, Christianity and Cultures.

\section*{Data Availability}

The ACT DR5 cluster catalog used in this paper is available on the NASA Legacy Archive Microwave Background Data Analysis (LAMBDA) website (\url{https://lambda.gsfc.nasa.gov}). MUSTANG-2 maps of individual clusters in this paper are on the Harvard Dataverse (\url{https://dataverse.harvard.edu/}).  The redMaPPer DR8 catalog can be downloaded from \url{http://risa.stanford.edu/redmapper/} while the MCXC catalog can be found at \url{https://heasarc.gsfc.nasa.gov/W3Browse/rosat/mcxc.html}.

\bibliographystyle{mnras}
\bibliography{point_src}

\begin{landscape}
\begin{table}
	\caption{Sources, flux densities and clusters. Cross IDs are found with a search radius of the largest of 9~arcsec or the resolution of the survey. The 1.4~GHz flux densities are from FIRST \citep{FIRST} except those marked $^a$ which are from NVSS \citep{Condon1998}; 28.5GHz data are from \citet{Coble2007}; The sources in Zwicky 3146 marked with a $^b$ are extended in the MUSTANG2 maps but are marked as seperate sources in FIRST; in this paper we combine the FIRST fluxes.  8.5 and 4.9~GHz data are from \citet{Lin2009}; the 74~MHz data is from the VLA Low-frequency Sky Survey \citep{Cohen2007}. An expanded machine readable version of this table is available as supplementary material. }
	\label{tab:src}
\begin{tabular}{lllllllllllr}
\hline
 & src RA & src Dec & S74~MHz  & S1.4~GHz & S4.9~GHz & S8.5~GHz & S28.5 GHz & S90 GHz & S500$\mu$m & WISE $3.4 \mu$m & $\alpha_{\mbox{{\tiny 1.4-90GHz}}}$  \\ 
Cluster     &  J2000 & J2000 & mJy  & mJy & mJy  & mJy & mJy & mJy & mJy     &  mJy \\ \hline
MOO J0105+1323 
 & 01:05:34.2 & 13:23:07.4 & --  & $<2.40 $ $^a$ & --  & --  & --  & $0.60\pm 0.07 $ & --   & $0.169\pm 0.007 $  & 0.221 \\
ACT-CL J0326-0043 
 & 03:26:50.1 & -00:43:49.4 & --  & $1.39\pm 0.14 $  & --  & --  & --  & $0.94\pm 0.10 $ & --   & $0.661\pm 0.018 $  & -0.093 \\
MOO J0448-1705 
 & 04:48:37.8 & -17:05:37.3 & --  & $4.30\pm 0.24 $ $^a$ & --  & --  & --  & $0.38\pm 0.05 $ & --   & $0.989\pm 0.021 $  & -0.583 \\
 & 04:48:42.2 & -17:04:55.5 & --  & $<2.40 $ $^a$ & --  & --  & --  & $0.37\pm 0.05 $ & --  & $<0.021 $  & 0.053 \\
MACS J0717.5+3745 
 & 07:17:37.0 & 37:44:20.4 & --  & $6.46\pm 0.14 $  & --  & --  & $3.29\pm 0.19 $  & $1.58\pm 0.16 $ & $<5.20 $   & $0.914\pm 0.035 $  & -0.338 \\
 & 07:17:38.0 & 37:46:49.0 & --  & $1.37\pm 0.15 $  & --  & --  & --  & $0.25\pm 0.04 $ & $<5.20 $   & $0.197\pm 0.009 $  & -0.413 \\
MOO J1001+6619 
 & 10:01:28.5 & 66:19:26.6 & --  & $7.38\pm 0.24 $ $^a$ & --  & --  & --  & $0.22\pm 0.03 $ & --   & $0.096\pm 0.016 $  & -0.842 \\
Zwicky 3146
 & 10:23:39.7 & 04:11:10.8 & --  & $2.04\pm 0.15 $  & $1.42\pm 0.07 $  & $0.98\pm 0.03 $  & $0.41\pm 0.07 $  & $0.19\pm 0.03 $ & $<6.20 $   & $0.925\pm 0.024 $  & -0.569 \\
 & 10:23:38.7 & 04:11:05.0 & --  & $<1.00 $ & $0.31\pm 0.02 $  & $0.37\pm 0.02 $  & --  & $0.05\pm 0.01 $ & $<6.20 $   & $0.286\pm 0.011 $  & -0.280 \\
 & 10:23:42.3 & 04:11:03.0 & --  & $<1.00 $ & --  & --  & --  & $0.04\pm 0.01 $ & $22.00\pm 6.20 $  & --  & -0.351 \\
 & 10:23:45.3 & 04:10:42.8$^b$ & --  & $89.21\pm 0.15 $  & --  & --  & $5.70\pm 0.31 $  & $1.51\pm 0.15 $ & $<6.20 $   & $0.114\pm 0.007 $  & -0.979 \\
 & 10:23:37.5 & 04:09:13.2$^b$ & --  & $27.13\pm 0.15 $  & --  & --  & $2.10\pm 0.16 $  & $0.81\pm 0.08 $ & $<6.20 $  & --  & -0.843 \\
 & 10:23:45.3 & 04:11:41.0 & --  & $6.06\pm 0.25 $  & --  & --  & $0.85\pm 0.10 $  & $0.18\pm 0.02 $ & $<6.20 $   & $0.130\pm 0.008 $  & -0.849 \\
MOO J1052+0823 
 & 10:52:14.1 & 08:24:53.5 & --  & $7.48\pm 0.14 $  & --  & --  & --  & $0.50\pm 0.05 $ & --   & $0.480\pm 0.026 $  & -0.647 \\
MOO J1054+0505 
 & 10:54:59.3 & 05:01:09.0 & --  & $5.81\pm 0.14 $  & --  & --  & --  & $4.75\pm 0.48 $ & --   & $0.066\pm 0.006 $  & -0.048 \\
 & 10:54:40.6 & 05:07:34.2 & --  & $3.56\pm 0.15 $  & --  & --  & --  & $0.58\pm 0.10 $ & --   & $0.081\pm 0.007 $  & -0.438 \\
MOO J1108+3242 
 & 11:08:53.4 & 32:45:03.7 & --  & $2.63\pm 0.11 $  & --  & --  & --  & $0.21\pm 0.04 $ & --   & $0.129\pm 0.006 $  & -0.604 \\
MOO J1110+6838 
 & 11:11:14.1 & 68:38:49.1 & --  & $<2.40 $ $^a$ & --  & --  & --  & $0.25\pm 0.04 $ & --   & $0.170\pm 0.006 $  & 0.008 \\
MOO J1142+1527 
 & 11:42:47.5 & 15:27:11.3 & --  & $37.17\pm 0.15 $  & --  & --  & --  & $0.67\pm 0.07 $ & --   & $0.199\pm 0.008 $  & -0.965 \\
MACS J1149.5+2223 
 & 11:49:22.5 & 22:23:25.3 & --  & $6.23\pm 0.15 $  & --  & --  & $3.40\pm 0.18 $  & $1.50\pm 0.16 $ & $<6.20 $   & $0.973\pm 0.024 $  & -0.342 \\
MOO J1322-0228 
 & 13:23:06.0 & -02:27:19.6 & --  & $<1.00 $ & --  & --  & --  & $0.41\pm 0.06 $ & --   & $0.324\pm 0.010 $  & 0.240 \\
RX J1347.5-1145 
 & 13:47:30.6 & -11:45:10.2 & $1006.00\pm 160.00 $  & $43.60\pm 0.24 $ $^a$ & --  & --  & $9.93\pm 0.23 $  & $3.31\pm 0.34 $ & $<9.00 $   & $0.623\pm 0.016 $  & -0.619 \\
MOO J1354+1329 
 & 13:54:55.5 & 13:29:34.0 & --  & $8.93\pm 0.15 $  & --  & --  & --  & $0.34\pm 0.04 $ & --   & $0.360\pm 0.010 $  & -0.788 \\
MOO J1506+5136 
 & 15:06:12.7 & 51:37:07.8 & --  & $6.23\pm 0.13 $  & --  & --  & --  & $2.75\pm 0.28 $ & --   & $0.349\pm 0.009 $  & -0.196 \\
 & 15:06:25.3 & 51:36:51.3 & --  & $35.16\pm 0.13 $  & --  & --  & --  & $0.88\pm 0.09 $ & --   & $0.149\pm 0.005 $  & -0.885 \\
 & 15:06:20.2 & 51:36:53.7 & --  & $8.56\pm 0.13 $  & --  & --  & --  & $0.51\pm 0.06 $ & --   & $0.091\pm 0.004 $  & -0.676 \\
 & 15:05:55.7 & 51:36:23.8 & --  & $10.16\pm 0.13 $  & --  & --  & --  & $1.05\pm 0.13 $ & --   & $0.030\pm 0.003 $  & -0.546 \\
Abell 2052 
 & 15:16:44.7 & 07:01:11.2 & --  & $3509.86\pm 0.38 $  & $897.60\pm 2.90 $  & $691.00\pm 2.70 $  & --  & $52.53\pm 5.25 $ & $27.80\pm 5.90 $   & $8.720\pm 0.185 $  & -1.009 \\
MOO J1554-0447 
 & 15:54:14.3 & -04:46:48.7 & $1310.00\pm 170.00 $  & $89.03\pm 0.24 $ $^a$ & --  & --  & --  & $6.82\pm 0.69 $ & --   & $0.255\pm 0.010 $  & -0.617 \\
 & 15:53:59.1 & -04:47:49.6 & --  & $<2.40$ $^a$ & --  & --  & --  & $0.56\pm 0.09 $ & --   & $0.226\pm 0.009 $  & 0.205 \\
\hline
	\end{tabular}   
\end{table}
\end{landscape}





\label{lastpage}
\end{document}